\documentclass[a4paper,fleqn,usenatbib,useAMS]{mnras}

\usepackage{graphicx}	% Including figure files
\usepackage{amsmath}	% Advanced maths commands
\usepackage{amssymb}	% Extra maths symbols
\usepackage{multicol}        % Multi-column entries in tables
\usepackage{bm}		% Bold maths symbols, including upright Greek
\usepackage{pdflscape}	% Landscape pagess
\usepackage[encapsulated]{CJK}
\usepackage{ucs}
\usepackage[utf8x]{inputenc}
\usepackage{multirow}
\usepackage{caption}
\usepackage{booktabs}
\usepackage{caption}
\usepackage[flushleft]{threeparttable}
\usepackage[dvipsnames]{xcolor}
\usepackage[shortlabels]{enumitem}
\usepackage{rotating}

\newcommand{\ha}{H$\alpha$ }
\newcommand{\hb}{H$\beta$ }

\usepackage[dvipsnames]{xcolor}
\usepackage{tikz,hyperref}

\definecolor{lime}{HTML}{A6CE39}
\DeclareRobustCommand{\orcidicon}{
	\begin{tikzpicture}
	\draw[lime, fill=lime] (0,0) 
	circle [radius=0.13] 
	node[white] {{\fontfamily{qag}\selectfont \tiny ID}};
	\draw[white, fill=white] (-0.0625,0.095) 
	circle [radius=0.007];
	\end{tikzpicture}
	\hspace{-2mm}
}

\foreach \x in {A, ..., Z}{\expandafter\xdef\csname orcid\x\endcsname{\noexpand\href{https://orcid.org/\csname orcidauthor\x\endcsname}
			{\noexpand\orcidicon}}
}
\usepackage[T1]{fontenc}
\usepackage{ae,aecompl}

\usepackage{newtxtext,newtxmath}
\title[The born-again planetary nebula A\,78]{Spatially-resolved spectroscopic investigation of the born-again planetary nebula A\,78}

\author[B.~Montoro-Molina, M.~A.~Guerrero \& J.~A.~Toal\'{a}]{B.~Montoro-Molina$^{1\orcidA}$\thanks{E-mail: borjamm@iaa.es},  M.~A.\,Guerrero$^{1\orcidB}$ and J.~A.\,Toal\'{a}$^{2\orcidC}\thanks{Visiting astronomer at the IAA-CSIC as part of the Centro de Excelencia Severo Ochoa Visiting-Incoming
program.}$\\
$^1$Instituto de Astrof\'\i sica de Andaluc\'\i a, IAA-CSIC, Glorieta de la Astronom\'\i a s/n, 18008, Granada, Spain\\
$^2$Instituto de Radioastronom\'{i}a y Astrof\'{i}sica, UNAM, Ant. Carretera a P\'{a}tzcuaro 8701, Ex-Hda. San Jos\'{e} de la Huerta, Morelia 58089, Mich., Mexico
}

\pubyear{2022}

\begin{document}
\label{firstpage}
\pagerange{\pageref{firstpage}--\pageref{lastpage}}
\maketitle

\begin{abstract}
We present the analysis of spatially-resolved spectroscopic observations of the born-again planetary nebula (PN) A\,78 that are used to investigate the chemistry and physical properties of its three main morphological components, namely the inner knots, its eye-like structure, and the low surface-brightness outer nebula. 

The H-poor chemical abundances of the inner knots confirm the born-again nature of A\,78, with a N/O abundances ratio consistent with the predictions of very late thermal pulses (VLTP).
On the other hand, the high Ne/O is not expected in VLTP events, which prompts a possible different evolutionary path may be involving a binary system.
The low N/O ratio and He/H abundances of the outer shell are indicative of a low-mass progenitor, whereas the chemical abundances of the eye-like structure, which results from the interaction between the H-poor born-again material and the outer nebula, evidence their mixture. 
Unlike previous works, the extinction is found to be inhomogeneous, being much higher towards the H-poor inner knots, where the presence of large amounts of C-rich dust has been previously reported. Dust-rich material seems to diffuse into outer nebular regions, resulting in zones of enhanced extinction. 

\end{abstract}


%%%%%%%%%%%%%%%%%%%%%%%%%%%%%%%%%%%%%%%%%%%%%%%%%%

%%%%%%%%%%%%%%%%% BODY OF PAPER %%%%%%%%%%%%%%%%%%

\section{Introduction}
\label{sec:intro}

The planetary nebula (PN) A\,78 consists of an old, elliptical nebula of low surface brightness surrounding a collection of clumps and filaments ordered in an eye-like structure and a series of central cometary knots distributed along a disk-like structure surrounding the central star (CSPN). 
These unusual structures inside the old PN of A\,78 were first imaged by \citet{Jacoby1979},  
although the limited spatial resolution of these images did not allow resolving in detail the structures close to the CSPN.
Spectroscopic observations revealed the innermost knots to be extremely H-deficient \citep{Jacoby1983}, leading to the speculation that they originated from a recent ejection of material inside the old PN by an unknown mechanism.

\citet{Iben1983} described a scenario where such ejections occurred in some PNe during the post-AGB phase of their CSPNe \citep[see also][]{Schonberner1979}, as they build a He shell with the critical mass to ignite into C and O and produce a very late thermal pulse (VLTP), ejecting highly processed, H-poor material inside the old H-rich PN. 
In a short period of time after this event \citep[$\sim$20--100~yr; e.g.,][]{Miller2006}, the star will cool as its envelope expands, returning it to the beginning of the post-ABG phase to later evolve for a second time into the white dwarf phase, developing a new fast stellar wind and increasing its ionizing photon flux. 

This event is rare and only a handful of objects have been suggested to have experienced it, including A\,30, A\,58, A\,78, FG~Sge (Hen\,1-5), V4334\,Sgr (the Sakurai's Object), HuBi\,1, and the nebula associated to WR\,72 \citep[see][and references therein]{Guerrero_etal2018,Gvaramadze2020,Ohnaka2022}. 
These are currently known as born-again PNe.

\citet{Clegg1993} were the first to resolve the innermost structures of A\,78, but the unprecedented capabilities of the {\it Hubble Space Telescope (HST)} disclosed in much great detail them \citep[][]{Borkowski1993}, revealing that the inner clumps and filaments appear to be distributed in a disrupted disk plus a pair of jet-like ejections. 
\citet{Meaburn1998} used high-resolution echelle spectra to show that the clumps in the disrupted disk expand with velocities of $\approx$25~km~s$^{-1}$, but the rest of the H-deficient structures in the eye-like structure reach velocities as high as $\sim$400~km~s$^{-1}$, in particular towards the Southwest of the CSPN.

Multi-epoch {\it HST} observations corroborated that structures closer to the CSPN expand slower than those located farther away \citep{Fang2014}. 
In addition, radiation-hydrodynamic simulations presented by those authors showed that the interaction of the current fast wind with the ejecta of a single, instantaneous VLTP event is able to reproduce the complex kinematics of born-again PNe such as A\,30 and A\,78. 
\citet{Fang2014} concluded that A\,78 experienced a born-again event approximately 1000 years ago. 
 
Since then, the CSPN of A78, which is now classified as a [WC] star, has increased its effective temperature up to 120~kK whilst developing a monstrous stellar wind with terminal velocity  $\gtrsim$3200 km~s$^{-1}$ \citep[][]{Heap1979,HB2004,Kaler&Feidelman1984,Toala2015}.

The born-again nature of A\,78 relies on the H-deficient chemical abundances of the recent ejecta, but very few spectroscopic analyses have been obtained up to now. 
\citet{Jacoby1983} presented the first intermediate-dispersion spectrum of A\,78 and determined chemical abundances on an aperture $\simeq$5$''$ Northeast of the CSPN.  
The authors highlighted the multiple-shell nature of this nebula and the scarcity of H and extremely enhanced He abundances, whereas the relative abundances of N, O and Ne were found to be typical of normal PNe. 
Accordingly it was concluded that the material is not chemically enriched, but rather the result of nuclear processing in a H-burning shell, with conversion to He approaching to 100\% completion.
Later on \citet{Manchado1988} presented the first spatially-resolved optical spectroscopic study of A\,78, including information on the eye-like and inner disk-like structures. 
The latter exhibits greatly enhanced chemical abundances not only of He, but also N, O and Ne, thus confirming the H-poor nature and chemical enrichment of the recent ejecta.  
The eye-like structure has also enhanced O and Ne abundances, but typical He and N abundances of PNe.  
The revision of the spectroscopic data of \citet{Jacoby1983} performed by \citet{Manchado1988} seems to imply that they probe a region mixing information from the eye-like and inner disk-like structures, with intermediate abundances. 
Finally, \citet{PCK2005} presented spectroscopic information of the eye-like of A\,78, but did not carry an extensive analysis.

\begin{figure*}
\centering
\includegraphics[width=0.49\linewidth]{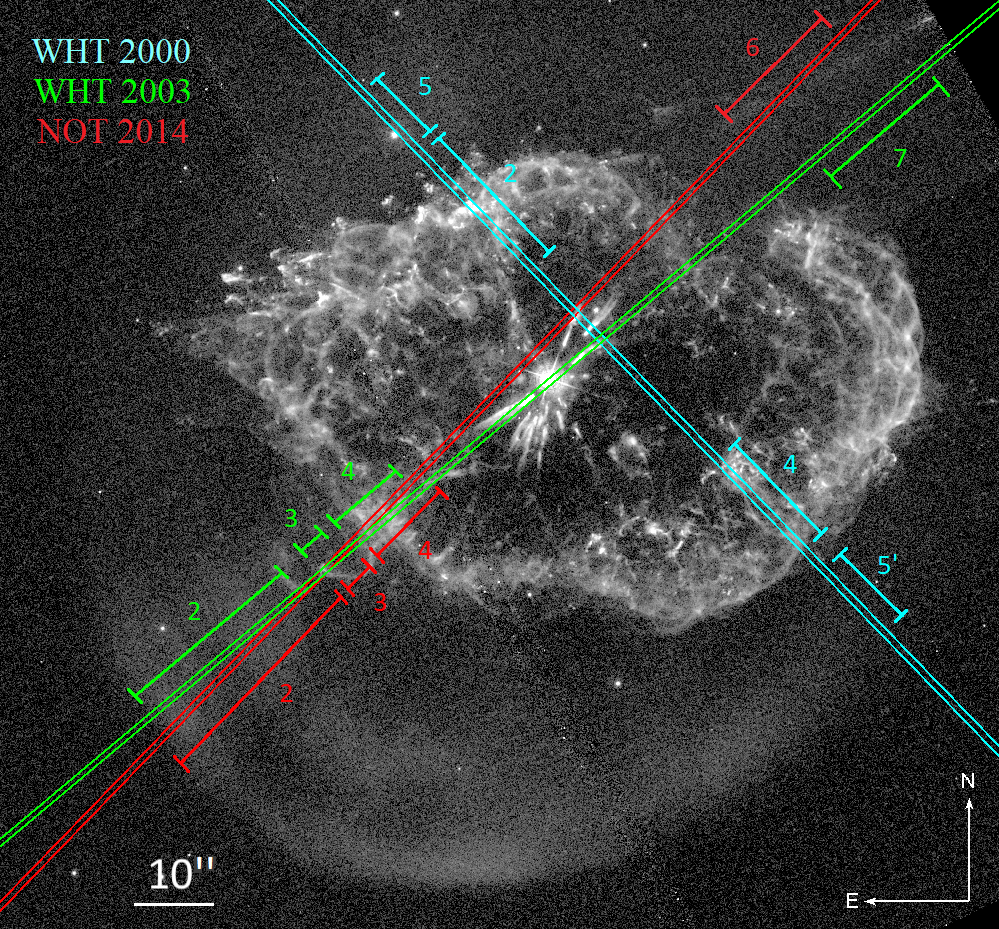} 
\includegraphics[width=0.49\linewidth]{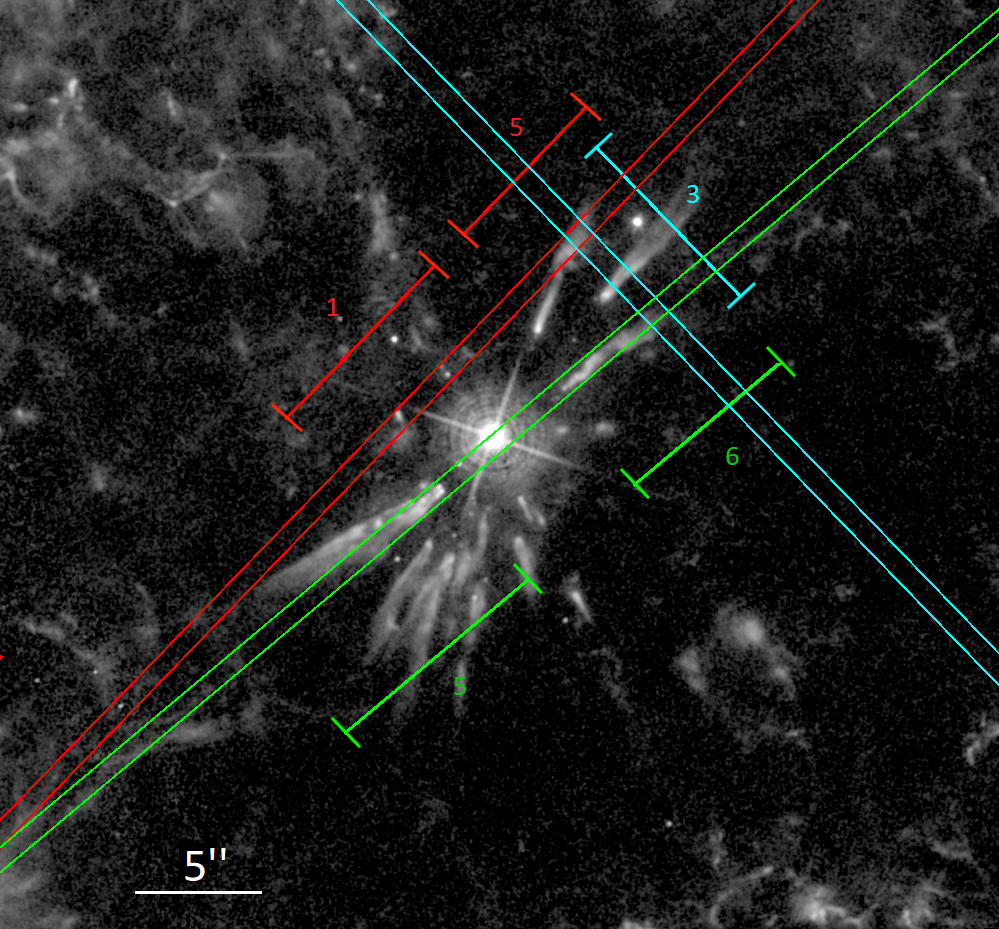}
\caption{
{\it HST} WFC3 F502N image of A\,78 (left), zooming into its innermost region (right). 
The slit positions covered at each epoch are marked with cyan (WHT 2000.50, PA=44$^\circ$), green (WHT 2003.58, PA=130$^\circ$) and red (NOT 2014.55, PA=136$^\circ$) rectangles, respectively.  
The spatial extent of the apertures used for spectra extraction are also marked with numbered segments of the same colours. 
}
\label{fig:A78_slit}
\end{figure*}

The previous works were mostly hampered by the limited quality of the pre-{\it HST} optical images available at their time, which made not possible to reliably assign the spectra to specific morphological features (e.g., a central cometary knot), thus making difficult their interpretation. 
On the other hand, the spectra mostly probed the bright eye-like structure, but no the outer nebula. 

In this work we present an analysis of optical long-slit intermediate-dispersion spectra of A\,78 including the outer nebula, the eye-like structure, and a number of central cometary knots, computing physical properties and abundances for them. 
Archival mid-infrarred Spitzer data is also used. 
The paper is organized as follows. 
The spectral observations and data reduction are described in Section~\ref{sec:obs}. 
Estimated parameters such as extinctions, temperatures, densities and subsequent ionic and total abundances are presented in Section~\ref{sec:results}. 
We discuss the results in Section~\ref{sec:diss} and finally present a summary in Section~\ref{sec:summary}.

\section{Observations and data preparation}
\label{sec:obs}
%\section{Observations and data ¿analysis?}

\subsection{Long-slit Optical Spectroscopy}

We have gathered optical long-slit intermediate-dispersion spectroscopic observations of A\,78 from three different epochs spanning from 2000 to 2014. 
These include two unpublished data sets available at the Isaac Newton Group (ING) archive\footnote{\url{http://casu.ast.cam.ac.uk/casuadc/ingarch/query}} obtained with the double-arm Intermediate-dispersion Spectrograph and Imaging System (ISIS) mounted on the 4.2m William Herschel Telescope (WHT) at the Observatorio del Roque de los Muchachos (ORM) on 2000 July 3 (2000.50) and 2003 August 1 (2003.58) 
and a long-slit intermediate-dispersion spectroscopic
observation obtained on 2014 July 20 (2014.55) with the ALhambra Faint Object Spectrograph and Camera (ALFOSC) mounted on the 2.5m Nordic Optical Telescope (NOT) of the ORM.

The total exposure times of the WHT ISIS observations were 4800 s (2$\times$2400 s) in 2000.50 and 16200 s (9$\times$1800 s) in 2003.58.
The R600B grating and EEV12 detector were used for the blue-arm in both epochs, providing a spatial scale of 0.19~arcsec~pix$^{-1}$ and a resolution $\sim$ 2530 at 4272 \AA.

As for the red-arm, the R316R grating was used in both epochs, but the TEK4 and MARCONI2 detectors on 2000.50 and 2003.58, respectively.  
The TEK4 detector provided in 2000.50 a spatial scale of 0.33~arcsec~pix$^{-1}$ and a resolution of 2430 at 6599 \AA

whereas the smaller pixel of the MARCONI2 detector provided in 2003.58 a spatial scale of 0.2~arcsec~pix$^{-1}$ and a resolution of 2460 at 6599 \AA.

The slit position during the 2003.58 observation was aligned along position angle (PA) of 130$^{\circ}$ such that the CSPN, a high-velocity component towards the Southeast of the eye-like structure, and the central knots were registered by the slit (see Fig.~\ref{fig:A78_slit}). 
The slit used in 2000.50 has a PA of 44$^{\circ}$, but it did not go through the CSPN.  
The lack of a clear fiducial reference made unclear the exact location of the slit, that we have otherwise determined comparing the emission profile in the [O~{\sc iii}] $\lambda$5007 emission line with the profile extracted from the {\it HST} WFC3 F502N image taken on 2012 November 22. The approximate slit positions are shown in Figure~\ref{fig:A78_slit}. 
The slit widths were 0.88 arcsec on 2000 and 0.76 arcsec on 2003 
(see Table \ref{tbl:observation}).

The NOT ALFOSC observations used the E2V CCD detector, providing a spatial scale of 0.21~arcsec~pix$^{-1}$. 
The slit position has an alignment quite similar to that of the 2003.58 observation, with a PA of 136$^{\circ}$, just slightly displaced from the CSPN to observe knots located along the polar direction of the inner disk 
(see Fig.~\ref{fig:A78_slit}). 
The slit width was set to 0.75 arcsec. 
Two 1200 s exposures were taken with the grisms \#7 and \#14, which have a resolution of 930 at 5260 \AA\ and 860 at 4630 \AA, respectively.

The spectral range 3189.0 to 6875.2 \AA\ was covered by these observations 
(see Table \ref{tbl:observation}). \\

\begin{table*}
\begin{center}
\caption{Journal of high-dispersion spectroscopic observations of A\,78.}
\label{tbl:observation}
% \centering
\begin{tabular}{lllccrcrlc} 
\hline
Telescope & Instrument & Detector &  Grating/Grism & $\lambda_c$ & Resolution & $\Delta\lambda$ & PA   & Date & Exp.\ Time  \\
          &            &          &                &       (\AA)        &            &  (\AA~pix$^{-1}$)      & ($^{\circ}$) &  &    (s)     \\
\hline
WHT 4.2m & ISIS     &  EEV12      & R600B  &  4272  &   2520~~~~    &   0.49 &  44  &  2000 July 4 & 2x2400~~~  \\
         &          &  TEK4       & R316R  &  6599 &   2430~~~~    &   1.47 &  44  &  2000 July 4 & 2x2400~~~  \\
         &          &  EEV12      & R600B  &  4272 &   2540~~~~    &   0.49 &  130 &  2003 August 3 & 9x1800~~~  \\
\smallskip
         &          &  MARCONI2   & R316R  &  6599 &   2460~~~~    &   0.82 &  130 &  2003 August 3 & 9x1800~~~  \\
NOT 2.5m  & ALFOSC  &  E2V CCD     & \#7     &  5260 &   930~~~~    &   1.49 &  136 &  2014 July 20 & 2x1200~~~  \\
          &         &  E2V CCD     & \#14    &  4630 &   860~~~~    &   1.40 &  136 &  2014 July 20 & 2x1200~~~  \\
\hline
\end{tabular}
\end{center}
\end{table*}

All data were processed following standard {\sc iraf} routines \citep{Tody1993}, which include bias subtraction and flat-field correction.

The wavelength calibration was carried out using ThAr and Ne arc lamps. The sky background was subsequently subtracted. The flux calibration was obtained using observations of spectro-photometric standard stars obtained in the same night as the object. We note that the slope of the continuum of the flux-calibrated spectrum of the standard star Hz\,44 extracted from the 2000.50 observations using the R316R grating is steeper than the available calibrations, most likely implying a calibration inconsistency. In this case, we applied an additional $\lambda$ dependent polynomial correction to make the flux-calibrated spectrum of the standard star consistent with the available calibrations.

\subsubsection{Comparing Multi-epoch Spectroscopic Observations}

In this work we will analyse multi-epoch long-slit spectroscopic observations of A\,78 obtained on 2000.50, 2003.58 and 2014.55, i.e., up to 14 years apart in time.  
Recent works on born-again PNe have revealed spectroscopic variations that are noticeable on short-time scales:  
the H-deficient ejecta of A\,58 has experienced a considerable brightening in the time from 1996 to 2021, with notable changes  in many emission line ratios \citep{M-M_etal2022}; 
the old outer nebula of HuBi\,1 is recombining as the result of the brightness decline of its central star in the last $\approx$50 years \citep{Guerrero_etal2018}; 
the H-rich old outer PN around Sakurai's Star (V4334\,Sgr) is also recombining, with line strength variations up to 3\% per year in the period 2007 to 2022 \citep{Reichel_etal2022}.  

We note that the central stars of these born-again PNe have luminosities and effective temperatures still varying in short-time scales, whereas those of A\,78 and its twin A\,30 have already entered a phase of stable luminosity and effective temperature. 
We will therefore assume that no noticeable spectral variations, neither in the inner VLTP ejecta nor in the outer nebular regions, have occurred in the time period from 2000 to 2014 considered here.

The comparison of multi-epoch spectroscopic data requires an accurate knowledge of the observation conditions (seeing, sky transparency, airmass, ...) and instrumental configuration (slit location and position angle, width, grating and its spectral dispersion, spatial scale, ...)  at each epoch.  
In order verify that the calibrations applied to the spectra obtained at the different epochs used here, we have compared  surface brightness profiles of the [O {\sc iii}] emission line, the brightest in A\,78, extracted from each spectrum to similar profiles extracted from the HST WFC3 F502N image.  
These are shown in Figure~\ref{fig:A78_profiles} in Appendix~\ref{apex:A}.  
The comparison of the profiles shape and surface brightness indicates that the flux calibrations of the spectroscopic data are consistent within 10 per cent.

\subsubsection{One-dimension Spectra Extraction}

The born-again PN A\,78 is composed of three main parts: a hydrogen-rich old outer nebula with an elliptical shape, 
a set of H-poor $\approx$1,000 yr old bright inner knots, and an intermediate eye-like structure.  
One-dimensional spectra have been extracted for these different structural components.  
The eye-like structure results from complex hydrodynamic interactions between the current fast wind from the CSPN, the previously ejected VLTP material, and the old outer nebula.  
This interaction produces high-speed flows as the result of the mass-load of the current stellar wind \citep{Meaburn1998}, but the projection of these features onto bright emission makes difficult its study.  
Interestingly, \citet{Fang2014} reported a high-velocity filament  (their ``Q'' and ``S'' features) towards the Southwest of the old outer nebula with a large tangential velocity in the plane of the sky of $\approx$600 km~s$^{-1}$ that is registered by the WHT 2003.58 and NOT 2014.55 observations.   
We have thus extracted 1D spectra for this component that will be hereafter referred as high-velocity filament (HVF).

The apertures used to extract 1D spectra are shown in Figure~\ref{fig:A78_slit} as numbered segments, using different colours according to the slit from which they have been extracted. 
In the following the apertures will be labelled according to the instrument used and observation year (W00, W03, and N14 refer to the WHT 2000, WHT 2003, and NOT 2014 data, respectively), adding the aperture number `X' as 'AX' to the label.  
Thus, for instance, aperture \#2 of the data obtained at the WHT in 2003 will be labelled as W03A2.
We note that the spectra of apertures W00A5 and W00A5$'$ on the outer nebula were added to increase the signal to noise ratio.  
This aperture will be denoted as W0055$'$.

The 1D spectra correspond to 
five apertures on the old nebula, 
four on the eye-like structure, 
two registering the HVF, 
four on central knots and its tails, and 
one on a central knot Northeast of the CSPN that presumably does not belong to the equatorial disk and has been claimed to be a polar knot.  
All the 1D spectra are shown in Appendix~\ref{apex:B}.

Lines in the spectra were identified and measured by fitting Gaussian profiles with the usual {\sc iraf} task \emph{splot}. 
A list of all the lines observed from the different extraction regions is given in Table \ref{tbl:lines_ratios_1}. Line fluxes are conventionally normalized to $F$(H$\beta$) = 100 in nebular studies, but because H$\beta$ is so weak in the innermost regions of A\,78, fluxes were instead normalized to $F$([O\,{\sc iii}]) = 100 to avoid large numbers.
We emphasize the high nebular excitation, with large [O~{\sc iii}] and He~{\sc ii} to H$\beta$ line ratios.  
The correspondence between the nebular excitation and the high effective temperature of its central star \citep[$>$110 kK;][]{HB2004,Toala2015} implies that the nebula is in equilibrium with the stellar ionizing flux, reassuring our assumption on the similar properties of spectra obtained 14 years apart.

\begin{table*}
\caption{
De-reddened emission line intensities of A\,78. 
Intensities are normalised to a value of 100 for the [O\,{\sc iii}] $\lambda$5007 \AA\ emission line.} 
\setlength{\tabcolsep}{0.5\tabcolsep}  
%\footnotesize
\begin{center}
\label{tbl:lines_ratios_1}
\begin{tabular}{l|ccccc}
\hline
          &   \multicolumn{5}{c|}{Old nebula}\\

\cmidrule(lr){2-6}
             & W03A2 & N14A2 & W00A55$^\prime$ & W03A7 & N14A6\\
\hline
$c$(H$\beta$)  & 0.40$\pm$0.07 & 0.37$\pm$0.03 & 0.03$\pm$0.00 & 0.15$\pm$0.02 & 0.14$\pm$0.06 \\
\hline

{{[O~{\sc ii}]} } 3726	&	2.90$\pm$1.97	&	$\dots$	&	$\dots$	&	$\dots$	&	$\dots$	\\
{{[O~{\sc ii}]} } 3729	&	4.27$\pm$1.98	&	$\dots$	&	$\dots$	&	$\dots$	&	$\dots$	\\
{{[Ne~{\sc iii}]} } 3868	&	11.82$\pm$1.47	&	7.01$\pm$0.89	&	7.89$\pm$0.92	&	13.82$\pm$3.47	&	$\dots$	\\
{H$\zeta$} 3889	&	$\dots$	&	$\dots$	&	4.19$\pm$1.09	&	$\dots$	&	$\dots$	\\
{{[Ne~{\sc iii}]} } 3967	&	4.02$\pm$1.38	&	$\dots$	&	$\dots$	&	5.23$\pm$3.19	&	$\dots$	\\
{H$\epsilon$} 3970	&	$\dots$	&	$\dots$	&	5.64$\pm$0.99	&	5.17$\pm$3.29	&	$\dots$	\\
{H$\delta$} 4101	&	3.41$\pm$0.99	&	$\dots$	&	11.75$\pm$1.06	&	9.26$\pm$3.29	&	$\dots$	\\
{H$\gamma$  } 4340	&	5.49$\pm$0.80	&	4.41$\pm$1.18	&	22.58$\pm$0.79	&	15.03$\pm$2.33	&	$\dots$	\\
{{[O~\sc{iii}]}   } 4363	&	$\dots$	&	$\dots$	&	2.41$\pm$0.67	&	3.73$\pm$2.33	&	$\dots$	\\
{He~\sc{ii} } 4686	&	12.32$\pm$0.73	&	11.85$\pm$0.45	&	67.80$\pm$1.03	&	38.23$\pm$3.13	&	39.32$\pm$1.89	\\
{H$\beta$   } 4861	&	11.87$\pm$0.68	&	10.48$\pm$0.43	&	49.80$\pm$0.93	&	29.22$\pm$2.70	&	31.96$\pm$1.55	\\
{{[O~\sc{iii}]}     } 4959	&	30.08$\pm$0.64	&	30.12$\pm$0.41	&	31.82$\pm$0.83	&	30.20$\pm$2.74	&	31.06$\pm$1.32	\\
{{[O~\sc{iii}]}    } 5007	&	100.00$\pm$0.88	&	100.00$\pm$0.38	&	100.00$\pm$1.30	&	100.00$\pm$3.00	&	100.00$\pm$1.95	\\
{H$\alpha$        } 6563	&	32.79$\pm$0.35	&	28.95$\pm$0.24	&	137.61$\pm$1.28	&	80.75$\pm$3.37	&	88.29$\pm$1.46	\\
{{[N~\sc{ii}]}     } 6584	&	1.40$\pm$0.27	&	$\dots$	&	$\dots$	&	$\dots$	&	$\dots$	\\
{{[S~\sc{ii}]}    } 6717	&	1.38$\pm$0.30	&	$\dots$	&	$\dots$	&	$\dots$	&	$\dots$	\\
{{[S~\sc{ii}]}  } 6731	&	1.18$\pm$0.37	&	$\dots$	&	$\dots$	&	$\dots$	&	$\dots$	\\
{{[Ar~\sc{iii}]}} 7135	&	1.63$\pm$0.54	&	$\dots$	&	$\dots$	&	$\dots$	&	
$\dots$	\\
{{[O~\sc{iv}]}} 25.9$\mu$m & 3.07$\pm$0.31   & 3.49$\pm$0.34 & $\dots$ & $\dots$ & $\dots$ \\
\hline
log~$F$([O~{\sc iii}]) \; (erg~cm$^{-2}$~s$^{-1}$)   & $-$14.026$\pm$0.003 & $-$14.010$\pm$0.001 & $-$14.733$\pm$0.003 & $-$14.358$\pm$0.009 & $-$14.497$\pm$0.006\\
\hline
$T_\mathrm{e}$([O~\sc{iii}]) \; (K) & $\dots$ & $\dots$ & 16700$^{+1100}_{-200}$ & 21200$^{+200}_{-700}$  & $\dots$\\

\hline
$n_\mathrm{e}$([O{~{\sc ii}}]) \; (cm$^{-3}$) & $<$ 100& $\dots$   & $\dots$ & $\dots$ & $\dots$ \\
$n_\mathrm{e}$([S~{\sc ii}])   \; (cm$^{-3}$)  & 370$_{-80}^{+100}$ & $\dots$ & $\dots$ & $\dots$ & $\dots$\\
\hline

\end{tabular}
\end{center}
\end{table*}

\subsection{Archival IR {\it Spitzer} observations}

After the analysis of the IR observations of the born-again PNe A\,30 and A\,78 presented by \citet{Toala2021}, we realised that the [O\,{\sc iv}] emission line at 25.9~\AA\, is one of the brightest emission lines present in the {\it Spitzer} Infrared Spectrograph (IRS) spectra. 
Thus, an appropriate O abundance determination should include this line in the calculations. 
For this, we retrieved from the NASA/IPAC Infrared Science Archive\footnote{\url{https://irsa.ipac.caltech.edu/frontpage/}} the {\it Spitzer} IRS dataset AOKey~10839808  obtained as part of the Program ID 3362 (PI: A.~Evans) on 2004 November 14. 
The map mode was used during the observations.  
The \emph{Spitzer} IRS observations of A\,78 used here cover its central regions as displayed in Figure~4-left of \citet{Toala2021}.

The {\it Spitzer} IRS spectra from the LL modules, which cover the 14--38.0~$\mu$m wavelength range including the [O\,{\sc iv}] 25.9~$\mu$m emission line, were processed making use of the CUbe Builder for IRS Spectra Maps \citep[{\sc cubism};][]{Smith2007}.  
One-dimensional spectra were extracted from the two-dimensional data cube for extraction regions identical to the W00A2, W00A4, W03A2, W03A3, W03A4, N14A2, N14A2, N14A3, and N14A4 apertures defined for the optical spectroscopic observations illustrated in Fig.~\ref{fig:A78_slit}.

\begin{table*}
\ContinuedFloat
\caption{
\textit{continued}. 
$^{\star}$: The NOT ALFOSC spectral resolution does not allow resolving the components of the [O~{\sc ii}] $\lambda\lambda$3726,3729 doublet and thus only the flux of the doublet is provided. }
\setlength{\tabcolsep}{0.5\tabcolsep}  
%\footnotesize
\begin{center}
\label{tbl:lines_ratios_1}
%\begin{tabular}{l|cc|cccc|cc}
\begin{tabular}{l|cccc|cc}
\hline
          &  \multicolumn{4}{c|}{Eye-like structure}  &  \multicolumn{2}{c|}{High-velocity filament} \\

\cmidrule(lr){2-5}\cmidrule(lr){6-7}
             & W00A2 & W00A4 & W03A4 & N14A4 & W03A3 & N14A3\\
\hline
$c$(H$\beta$)  & 0.17$\pm$0.03 & 0.10$\pm$0.03 & 0.24$\pm$0.07 & 0.13$\pm$0.04  & 0.52$\pm$0.1 & 0.26$\pm$0.09 \\
\hline

{{[O~{\sc ii}]} } 3726	&	0.92$\pm$0.47	&	4.30$\pm$0.24	&	6.59$\pm$0.93	&	$\dots$	&	5.18$\pm$2.24	&	$\dots$	\\
{{[O~{\sc ii}]} } 3729	&	0.96$\pm$0.54	&	3.61$\pm$0.25	&	6.98$\pm$0.80	&	9.59$\pm$0.44	&	9.89$\pm$2.63	&	\ 17.26$\pm$11.34$^{\star}$	\\
{{[Ne~{\sc iii}]} } 3868	&	14.00$\pm$0.26	&	16.54$\pm$0.16	&	14.71$\pm$0.54	&	16.11$\pm$1.60	&	15.98$\pm$1.65	&	12.86$\pm$10.85	\\
{He~{\sc i}} 3887 + {H$\zeta$} 3889	&	0.42$\pm$0.2	&	0.70$\pm$0.23	&	$\dots$	&	$\dots$	&	$\dots$	&	$\dots$	\\
{{[Ne~{\sc iii}]} } 3967	&	4.24$\pm$0.27	&	5.14$\pm$0.13	&	4.60$\pm$0.55	&	3.86$\pm$1.61	&	8.98$\pm$2.54	&	5.64$\pm$0.56	\\
{H$\epsilon$} 3970	&	0.56$\pm$0.24	&	0.59$\pm$0.13	&	$\dots$	&	$\dots$	&	$\dots$	&	$\dots$	\\
{H$\delta$} 4101	&	1.60$\pm$0.22	&	1.00$\pm$0.17	&	1.60$\pm$0.60	&	$\dots$	&	3.57$\pm$1.86	&	$\dots$	\\
{H$\gamma$  } 4340	&	3.27$\pm$0.22	&	2.04$\pm$0.14	&	2.02$\pm$0.32	&	2.24$\pm$0.24	&	4.36$\pm$0.88	&	5.83$\pm$0.73	\\
{{[O~\sc{iii}]}   } 4363	&	2.95$\pm$0.21	&	2.91$\pm$0.14	&	2.94$\pm$0.37	&	2.47$\pm$0.20	&	3.46$\pm$1.04	&	$\dots$	\\
{He~\sc{ii} } 4686	&	9.96$\pm$0.20	&	6.95$\pm$0.11	&	5.13$\pm$0.31	&	4.46$\pm$0.15	&	10.94$\pm$0.78	&	8.59$\pm$0.49	\\
{{[Ar~{\sc iv}]}} 4711	&	1.03$\pm$0.16	&	1.20$\pm$0.12	&	$\dots$	&	$\dots$	&	$\dots$	&	$\dots$	\\
{He~{\sc i}} 4713 + {{[Ne~{\sc iv}]}} 4715	&	0.71$\pm$0.20	&	0.66$\pm$0.11	&	$\dots$	&	$\dots$	&	$\dots$	&	$\dots$	\\
{{[Ne~{\sc iv}]}} 4724	&	0.99$\pm$0.21	&	1.22$\pm$0.13	&	$\dots$	&	$\dots$	&	$\dots$	&	$\dots$	\\
{{[Ar~{\sc iv}]}} 4740	&	0.93$\pm$0.16	&	0.74$\pm$0.10	&	$\dots$	&	$\dots$	&	$\dots$	&	$\dots$	\\
{H$\beta$   } 4861	&	6.80$\pm$0.16	&	4.27$\pm$0.09	&	4.79$\pm$0.27	&	4.14$\pm$0.15	&	9.07$\pm$0.69	&	8.22$\pm$0.54	\\
{{[O~\sc{iii}]}     } 4959	&	30.44$\pm$0.20	&	32.19$\pm$0.08	&	29.87$\pm$0.36	&	31.82$\pm$0.15	&	30.14$\pm$0.89	&	31.28$\pm$0.50	\\
{{[O~\sc{iii}]}    } 5007	&	100.00$\pm$0.04	&	100.00$\pm$0.11	&	100.00$\pm$0.47	&	100.00$\pm$0.20	&	100.00$\pm$1.18	&	100.00$\pm$0.68	\\
{{[O~{\sc i}]}  } 6300	&	$\dots$	&	0.60$\pm$0.04	&	$\dots$	&	$\dots$	&	$\dots$	&	$\dots$	\\
{{[N~\sc{ii}]}    } 6548	&	0.22$\pm$0.07	&	0.73$\pm$0.03	&	0.88$\pm$0.09	&	1.08$\pm$0.10	&	1.74$\pm$0.31	&	1.58$\pm$0.20	\\
{H$\alpha$        } 6563	&	18.70$\pm$0.04	&	11.78$\pm$0.03	&	13.20$\pm$0.09	&	11.43$\pm$0.09	&	24.85$\pm$0.36	&	22.70$\pm$0.26	\\
{{[N~\sc{ii}]}     } 6584	&	0.63$\pm$0.06	&	2.08$\pm$0.03	&	2.73$\pm$0.09	&	3.31$\pm$0.09	&	4.69$\pm$0.30	&	4.97$\pm$0.23	\\
{He~{\sc i}     } 6678	&	0.12$\pm$0.05	&	0.10$\pm$0.02	&	0.17$\pm$0.10	&	$\dots$	&	$\dots$	&	$\dots$	\\
{{[S~\sc{ii}]}    } 6717	&	0.14$\pm$0.04	&	0.08$\pm$0.02	&	0.51$\pm$0.08	&	$\dots$	&	1.06$\pm$0.39	&	$\dots$	\\
{{[S~\sc{ii}]}  } 6731	&	0.08$\pm$0.02	&	0.10$\pm$0.02	&	0.39$\pm$0.08	&	$\dots$	&	1.31$\pm$0.39	&	$\dots$	\\
{{[Ar~{\sc v}]}  } 7005	&	0.23$\pm$0.03	&	0.23$\pm$0.02	&	$\dots$	&	$\dots$	&	$\dots$	&	$\dots$	\\
{{[Ar~\sc{iii}]}} 7135	&	0.36$\pm$0.03	&	0.32$\pm$0.02	&	0.95$\pm$0.06	&	$\dots$	&	1.21$\pm$0.31	&	$\dots$	\\
%{{[O~\sc{ii}]}    } 7320	&	$\dots$	&	0.36$\pm$0.01	&	$\dots$	&	$\dots$	&	$\dots$	&	$\dots$	\\
%{{[O~\sc{ii}]}    } 7330	&	$\dots$	&	0.33$\pm$0.02	&	$\dots$	&	$\dots$	&	$\dots$	&	$\dots$	\\
{{[O~\sc{iv}]}} 25.9$\mu$m & 14.97$\pm$1.39   & 6.20$\pm$0.61 & 5.55$\pm$0.55 & 7.07$\pm$0.70 & 18.16$\pm$1.91 &  26.80$\pm$2.60 \\
\hline
log~$F$([O~{\sc iii}]) \; (erg~cm$^{-2}$~s$^{-1}$)  & $-$13.785$\pm$0.001 & $-$13.556$\pm$0.003 & $-$13.823$\pm$0.001 & $-$13.811$\pm$0.001  & $-$14.673$\pm$0.003 & $-$14.573$\pm$0.002 \\
\hline

$T_\mathrm{e}$([O~\sc{iii}]) \; (K)  & 18480$_{-800}^{+780}$ & 18480$_{-930}^{+530}$ & 17890$_{-170}^{+1300}$ & $\dots$   & 18370$_{-600}^{+1300}$ & $\dots$ \\

\hline
$n_\mathrm{e}$([O{~\sc{ii}}]) \; (cm$^{-3}$)    & 590$_{-210}^{+90}$     & 1110$_{-240}^{+210}$ & 500$_{-140} ^{+140}$ & $\dots$   & $\dots$                & $\dots$ \\
$n_\mathrm{e}$([S~{\sc ii}]) \; (cm$^{-3}$)    & $\dots$               & 1910$_{-440}^{+480}$ & 140$_{-130}^{+140}$  & $\dots$   & 1940$_{-520}^{+400}$   & $\dots$ \\
$n_\mathrm{e}$([Ar~{\sc iv}]) \; (cm$^{-3}$)   & 2300$\pm900$   & $\dots$               & $\dots$               & $\dots$   & 4500$_{-1720}^{+1760}$ & $\dots$ \\
\hline

\end{tabular}
\end{center}
\end{table*}

\section{Results}
\label{sec:results}

\subsection{Extinction}
\label{extinction}

We have measured the extinction values for each aperture described in the previous section using the typical Balmer decrement method based on the H$\alpha$/H$\beta$ ratio. 
It shall be noticed, however, that our spectra do not resolve the He~{\sc ii} Pickering series lines at 4860 \AA\ and 6560 \AA\ from H$\beta$ and H$\alpha$, respectively. 
Given the significant brightness of the He~{\sc ii} $\lambda$4686 line, the \ha and \hb lines can be expected to be notably contaminated from the He~{\sc ii} Pickering lines. 
In the most extreme cases of the H-poor internal knots, most of the emission initially attributed to \ha and \hb will actually correspond to emission from the adjacent He~{\sc ii} Pickering lines. 
An accurate estimate of this contribution is crucial because a reliable measurement of the \ha and \hb fluxes are needed for deriving the extinction, but also for the subsequent determination of abundances.

Since the theoretical H$\alpha$/H$\beta$, 
He~{\sc ii}~$\lambda$4860/He~{\sc ii}~$\lambda$4686, and 
He~{\sc ii}~$\lambda$6560/He~{\sc ii}~$\lambda$4686 line ratios depend on the electronic density $n_\mathrm{e}$ and temperature $T_\mathrm{e}$, a prior knowledge of the physical conditions is required to correct the contamination of the He~{\sc ii} Pickering to the H~{\sc i} Balmer lines. 
As for the density, the extinction correction has little effect, even for high extinction values, because the wavelength proximity of the density-sensitive diagnostic doublets.  
As for the temperature, however, the effects of these corrections acquire notable relevance. 
To solve this problem we have carried out an iterative process which consists of: 
1) computing initial values of $T_\mathrm{e}$ and $n_\mathrm{e}$ (the details of these calculations are given in next section), 
2) using those to derive the theoretical H$\alpha$/H$\beta$, 
He~{\sc ii}~$\lambda$4860/He~{\sc ii}~$\lambda$4686, and 
He~{\sc ii}~$\lambda$6560/He~{\sc ii}~$\lambda$4686 line ratios,
3) using these values to solve the equations system described in Appendix~\ref{apex:C} to provide us with a first approximation of the \ha and \hb fluxes free from the contamination of He~{\sc ii} lines, 
4) using the \ha and \hb fluxes to compute the extinction coefficient $c$(H$\beta$), and 
5) using this to de-redden the corresponding density and temperature sensitive emission lines to compute $T_\mathrm{e}$ and $n_\mathrm{e}$ and start a new iteration. 
This loop ends up when the difference between the new and previous values of $c$(H$\beta$) differs by less than 0.01.

The values of $c$(H$\beta$) derived for the different apertures are listed in Table~\ref{tbl:lines_ratios_1}.
After correcting from the contribution of the He~{\sc ii} Pickering lines to H$\alpha$ and H$\beta$, their fluxes in the apertures registering the H-poor knots W00A3, W03A5, and W03A6 are significantly reduced to 
% 2.28$\times 10^{-16}$, 
% 8.23$\times 10^{-17}$, and 
% 2.32$\times 10^{-16}$ 
2.3$\times 10^{-16}$, 
8.2$\times 10^{-17}$, and 
2.3$\times 10^{-16}$ 
erg~cm$^{-2}$~s$^{-1}$, respectively, for H$\alpha$, and   
% 8.33$\times 10^{-17}$, 
% 2.99$\times 10^{-17}$, and 
% 8.52$\times 10^{-17}$ 
8.3$\times 10^{-17}$, 
3.0$\times 10^{-17}$, and 
8.5$\times 10^{-17}$ 
erg~cm$^{-2}$~s$^{-1}$, respectively, for H$\beta$. 
The values of $c$(H$\beta$) for these apertures are 1.8$\pm$1.1, 1.3$\pm$0.6, and 1.6$\pm$1.0, where the large error-bars result from the significant uncertainties affecting the detection of the H~{\sc i} Balmer lines in these apertures.  
Lower limits of extinction $\geq$0.6 and $\geq$0.5 were set for apertures N14A5 and N14A1, where H$\beta$ is not detected.

\subsection{Physical conditions of the nebula}

We used {\sc PyNeb} \citep{pyneb2015} to obtain the physical conditions for each nebular aperture of A\,78. 
The electronic density was derived from the line ratios of the 
[O~{\sc ii}] $\lambda$3726/$\lambda$3728, 
[S~{\sc ii}] $\lambda$6716/$\lambda$6730, and 
[Ar~{\sc iv}] $\lambda$4711/$\lambda$4740 density-sensitive doublets, and the electronic temperature from the [O~{\sc iii}] $\lambda$4363/$\lambda\lambda$4959+5007 auroral-to-nebular line ratio. 
We note that {\sc PyNeb} computes simultaneously $T_\mathrm{e}$ and $n_\mathrm{e}$, including the cross-dependences of both parameters. 
The results are presented at the bottom lines of Table~\ref{tbl:lines_ratios_1}.
When available, the [O {\sc ii}] doublet and [O {\sc iii}] emission lines were adopted for the calculation of density and temperature, respectively. 
In apertures where it was not possible to measure the components of the [O {\sc ii}] doublet, the fainter [S {\sc ii}] and [Ar {\sc iv}] lines were used instead. 
The densities derived from these are deemed less reliable given the uncertainties in their fluxes.

The values listed in Table~\ref{tbl:lines_ratios_1} show that the average temperature in the eye-like structure and HVF can be estimated to be around 18400 K, which matches the temperatures derived by \citet{Manchado1988}, but it is higher than the value of 13000~K derived by \citet{Jacoby1983}. 
The central knots present slightly higher $T_\mathrm{e}$, reaching up to 23000 K, whereas the temperature estimates for the outer nebula range from 16700 to 21200~K. 

As for the density derived from the [O~{\sc ii}] doublet, its value ranges from 500 to 1000 cm$^{-3}$ for the eye-like structure, and from 110 (i.e., the lower limit) to 580 cm$^{-3}$ for the inner knots.  
The density of the outer nebula, as derived from the [S~{\sc ii}] doublet, is 370 cm$^{-3}$.  
These figures of density are generally higher than the low-density regime $\approx$100 cm$^{-3}$ previously reported \citep{Manchado1988}. 
For apertures where temperature or density could not be calculated, fixed values of 18000~K and 300~cm$^{-3}$ were adopted for the extinction and abundances calculations.

\begin{table*}
\ContinuedFloat
\footnotesize
\begin{center}
\caption{\textit{continued}}
\label{tbl:lines_ratios_2}
\begin{tabular}{l|ccccc|c}

\hline
            & \multicolumn{4}{c}{Disk knots and tails}  &  Polar knot\\
\cmidrule(lr){2-5}\cmidrule(lr){6-6}
            & W00A3 & W03A5 & W03A6 & N14A5 & N14A1  \\
\hline
% $c$(H$\beta$) & 1.78$\pm$1.06 & 1.30$\pm$0.61 & 1.61$\pm$0.96 & $>0.61$ & $>0.53$  \\
$c$(H$\beta$) & 1.8$\pm$1.1 & 1.3$\pm$0.6 & 1.6$\pm$1.0 & $\geq0.6$ & $\geq0.5$  \\
\hline

{{[Ne~{\sc v}]}} 3345	&	$\dots$	&	$\dots$	&	$\dots$	&	202.57$\pm$17.15	&	248.10$\pm$27.64	\\
{{[Ne~{\sc v}]}} 3426	&	313.80$\pm$10.58	&	182.88$\pm$12.63	&	430.91$\pm$49.70	&	432.82$\pm$15.82	&	706.69$\pm$22.61	\\
{{[O~{\sc ii}]} } 3726	&	10.10$\pm$4.95	&	5.60$\pm$1.04	&	8.65$\pm$1.96	&	$\dots$	&	$\dots$	\\
{{[O~{\sc ii}]} } 3729	&	7.47$\pm$0.47	&	4.91$\pm$1.03	&	8.34$\pm$2.17	&	8.06$\pm$1.44	&	$\dots$	\\
{{[Ne~{\sc iii}]} } 3868	&	41.65$\pm$0.36	&	26.15$\pm$0.71	&	35.93$\pm$1.44	&	42.19$\pm$1.43	&	33.95$\pm$1.87	\\
{He~{\sc i}} 3887	&	1.22$\pm$0.27	&	$\dots$	&	$\dots$	&	$\dots$	&	$\dots$	\\
{{[Ne~{\sc iii}]} } 3967	&	11.22$\pm$0.25	&	8.10$\pm$0.65	&	9.50$\pm$1.24	&	12.22$\pm$1.44	&	$\dots$	\\
{?} 4358	&	$\dots$	&	$\dots$	&	1.83$\pm$0.78	&	$\dots$	&	$\dots$	\\
{{[O~\sc{iii}]}   } 4363	&	3.38$\pm$0.16	&	3.03$\pm$0.35	&	4.22$\pm$0.79	&	3.79$\pm$1.21	&	6.45$\pm$1.21	\\
{He~\sc{i} } 4471	&	0.91$\pm$0.13	&	0.68$\pm$0.28	&	$\dots$	&	$\dots$	&	$\dots$	\\
{He~{\sc ii}   } 4543	&	$\dots$	&	0.78$\pm$0.20	&	$\dots$	&	$\dots$	&	$\dots$	\\
{{[Fe~{\sc iii}]}} 4658	&	$\dots$	&	1.61$\pm$0.35	&	$\dots$	&	$\dots$	&	$\dots$	\\
{He~\sc{ii} } 4686	&	10.79$\pm$0.10	&	11.89$\pm$0.25	&	13.09$\pm$0.55	&	8.76$\pm$0.34	&	14.56$\pm$0.97	\\
{{[Ne~{\sc iv}]}} 4715	&	4.64$\pm$0.12	&	4.36$\pm$0.29	&	5.79$\pm$0.69	&	4.93$\pm$0.34	&	11.41$\pm$0.97	\\
{{[Ne~{\sc iv}]}} 4724	&	7.57$\pm$0.13	&	7.25$\pm$0.30	&	8.92$\pm$0.70	&	7.07$\pm$0.37	&	14.85$\pm$0.96	\\
{{[Ar~{\sc iv}]}} 4740	&	$\dots$	&	0.77$\pm$0.32	&	0.87$\pm$0.57	&	$\dots$	&	$\dots$	\\
% {H$\beta$   } 4861	&	0.01$\pm$0.09*	&	0.00$\pm$0.32*	&	0.02$\pm$0.49*	&	$\dots$	&	$\dots$	\\
{H$\beta$   } 4861	&	0.0003:	&	0.001:	&	0.006: &	$\dots$	&	$\dots$	\\
{{[O~\sc{iii}]}     } 4959	&	33.79$\pm$0.07	&	31.91$\pm$0.29	&	32.62$\pm$0.43	&	33.57$\pm$0.24	&	32.84$\pm$0.45	\\
{{[O~\sc{iii}]}    } 5007	&	100.00$\pm$0.09	&	100.00$\pm$0.38	&	100.00$\pm$0.56	&	100.00$\pm$0.32	&	100.00$\pm$0.58	\\
{He~\sc{i} } 5876	&	1.77$\pm$0.03	&	1.29$\pm$0.04	&	1.26$\pm$0.69	&	1.20$\pm$0.09	&	$\dots$	\\
{{[O~{\sc i}]}  } 6300	&	0.34$\pm$0.02	&	0.73$\pm$0.05	&	0.61$\pm$0.07	&	$\dots$	&	$\dots$	\\
{{[O~{\sc i}]}} 6363	&	0.06$\pm$0.02	&	$\dots$	&	$\dots$	&	$\dots$	&	$\dots$	\\
{[Ar~{\sc v}]    } 6435	&	0.11$\pm$0.02	&	$\dots$	&	$\dots$	&	$\dots$	&	$\dots$	\\
{?} 6500	&	$\dots$	&	$\dots$	&	$\dots$	&	1.22$\pm$0.05	&	1.79$\pm$0.09	\\
{{[N~\sc{ii}]}    } 6548	&	0.34$\pm$0.01	&	0.78$\pm$0.03	&	0.35$\pm$0.04	&	$\dots$	&	$\dots$	\\
% {H$\alpha$        } 6563	&	0.04$\pm$0.02*	&	0.01$\pm$0.04*	&	0.06$\pm$0.04*	&	1.40$\pm$0.05	&	1.28$\pm$0.09	\\
{H$\alpha$        } 6563	&	0.04: &	0.01: & 0.06: &	1.40$\pm$0.05	&	1.28$\pm$0.09	\\
{{[N~\sc{ii}]}     } 6584	&	1.00$\pm$0.01	&	2.14$\pm$0.03	&	0.98$\pm$0.04	&	0.52$\pm$0.05	&	$\dots$	\\
{He~{\sc i}     } 6678	&	0.22$\pm$0.01	&	$\dots$	&	0.23$\pm$0.04	&	$\dots$	&	$\dots$	\\
{{[S~\sc{ii}]}    } 6717	&	$\dots$	&	0.07$\pm$0.05	&	$\dots$	&	$\dots$	&	$\dots$	\\
{{[S~\sc{ii}]}  } 6731	&	$\dots$	&	0.04$\pm$0.02	&	$\dots$	&	$\dots$	&	$\dots$	\\
{{[Ar~{\sc v}]}  } 7005	&	0.09$\pm$0.03	&	0.28$\pm$0.04	&	0.26$\pm$0.07	&	$\dots$	&	$\dots$	\\
{He~{\sc i}     } 7065	&	0.05$\pm$0.03	&	0.19$\pm$0.04	&	0.12$\pm$0.07	&	$\dots$	&	$\dots$	\\
{{[Ar~\sc{iii}]}} 7135	&	0.05$\pm$0.03	&	0.19$\pm$0.03	&	0.09$\pm$0.06	&	$\dots$	&	$\dots$	\\
%{{[O~\sc{ii}]}    } 7320	&	0.10$\pm$0.03	&	0.48$\pm$0.05	&	$\dots$	&	$\dots$	&	$\dots$	\\
%{{[O~\sc{ii}]}    } 7330	&	0.06$\pm$0.03	&	0.35$\pm$0.04	&	$\dots$	&	$\dots$	&	$\dots$	\\

\hline
log~$F$([O~{\sc iii}]) \; (erg~cm$^{-2}$~s$^{-1}$)  & $-$13.935$\pm$0.001 & $-$13.438$\pm$0.001  & $-$13.951$\pm$0.002 & $-$14.126$\pm$0.001 & $-$14.256$\pm$0.002 \\
\hline 

$T_\mathrm{e}$([O~\sc{iii}]) \; (K)  & 19870$_{-850}^{+1680}$ & 18750$_{-730}^{+730}$  & 23140$_{-1070}^{+1270}$ & $\dots$  & $\dots$ \\

\hline
$n_\mathrm{e}$([O{~\sc{ii}}]) \; (cm$^{-3}$)  & 110$_{-110}^{+240}$ & 370$_{-360}^{+240}$  & 580$_{-160}^{+190}$ & $\dots$  & $\dots$  \\
%n$_e$([S~\sc{ii}])   & $\dots$             & $\dots$              & $\dots$             & $\dots$  & $\dots$  \\
%n$_e$([Ar~\sc{iv}])  & $\dots$             & $\dots$              & $\dots$             & $\dots$  & $\dots$  \\
\hline

\end{tabular}
\end{center}
\end{table*}

\subsection{Chemical abundances}

In born-again PNe, the abundances of the outer old nebula are expected to be typical of PNe, but the VLTP ejecta is formed by H-poor material. 
Therefore, the calculation of abundances in born-again PNe, as is the case of A\,78, is far from simple because the H-deficient material of the inner ejecta is seen through the outer H-rich nebula. 
A precise determination of the fluxes of the \ha and \hb lines in the innermost regions of born-again PNe is cumbersome, but crucial to derive their proper chemical abundances \citep[e.g.,][]{MM2022a}.

Abundances relative to H for ionic species of O, N, Ne and S were derived from collisional excitation lines (CELs). 
The spectra are not deep enough to detect optical recombination lines (ORLs) for those elements, thus it was not possible to estimate the abundance discrepancy factor (ADF). 
This is expected to be high in the central knots, as it is the case for the inner knots of A\,30, which present the highest ADF values among PNe \citep{Wesson2003,Simpson2022}. 
The temperatures and densities adopted for abundances calculations were the same as for the extinction determination.

\begin{table}
\footnotesize
\begin{center}
\caption{List of references of the transition probabilities and collision strengths adopted to calculate ionic abundances. }
\setlength{\tabcolsep}{0.75\tabcolsep}    
\label{tbl:abn_references}
%\resizebox{\linewidth}{!}{
\begin{tabular}{lll}
%\vspace{0.15cm}
\hline
Ions    &   Transition probabilities    &   Collision strengths  \\
\hline
{{[O {\sc ii}]}}	&	\citet{Zeippen1982}       &	\citet{Kisielius2009}	 \\
{{[O~{\sc iii}]}} 	&	\citet{FroeseFischer2004} &	\citet{Storey2014}	\\
{{[N {\sc ii}]}}	&	\citet{FroeseFischer2004} &	\citet{Tayal2011}	\\
{{[Ne {\sc ii}]}}	&	\citet{Galavis1997}       &	\citet{McLaughlin2000} \\
{{[Ne {\sc iii}]}}	&	\citet{Godefroid1984}     &	\citet{Giles1981}  \\
{{[Ne {\sc iv}]}}	&	\citet{Galavis1997}       & \citet{Dance2013}	\\
{{[Ar {\sc iv}]}}	&	\citet{Rynkun2019}	      &	\citet{Ramsbottom1997} \\
{{[S {\sc ii}]}}	&	\citet{Rynkun2019}        &	\citet{Tayal2010}	\\
\hline
\end{tabular}
%}
\end{center}
\end{table}

\begin{table*}
\caption{Ionic abundances by number (12 + log\,$\frac{N(\mathrm{X^{+i})}}{N(\mathrm{H})}$) obtained from CELs. } 
\setlength{\tabcolsep}{0.8\tabcolsep}  
%\footnotesize
\begin{center}
\setlength{\tabcolsep}{0.5\tabcolsep}
\label{tbl:abn_CELs}
\begin{tabular}{l|c|ccccc|cccc|cc|ccc}
\hline
    &  &   
\multicolumn{5}{c}{Old nebula}  &  
\multicolumn{4}{c}{Eye-like structure}  &  
\multicolumn{2}{c}{High-velocity filament}  & 
\multicolumn{3}{c}{Disk knots and tails} \\
\cmidrule(lr){3-7}\cmidrule(lr){8-11}\cmidrule(lr){12-13}\cmidrule(lr){14-16}
        &    & W03A2 & N14A2 & W00A55$^\prime$ & W03A7 & N14A6 & W00A2 & W00A4 & W03A4 & N14A4 & W03A3 & N14A3 & W00A3 & W03A5 & W03A6 \\
\hline

O$^{+}$/H$^+$	&	$\lambda \lambda$ 3726, 3729	&	6.48	&	$\dots$	&	$\dots$	&	$\dots$	&	$\dots$	&	6.17	&	6.96	&	7.12	&	7.29	&	6.93	&	7.25	&	9.69	&	10.03	&	9.36	\\
O$^{2+}$/H$^+$	&	$\lambda \lambda$ 4959, 5007	&	7.73	&	7.78	&	7.20	&	7.21	&	7.30	&	8.00	&	8.21	&	8.13	&	8.22	&	7.87	&	7.92	&	10.62	&	11.14	&	10.28	\\
O$^{3+}$/H$^+$	&	$\lambda$ 25.9$\mu$m	&	6.52	&	6.60	&	$\dots$	&	$\dots$	&	$\dots$	&	7.46	&	7.32	&	7.18	&	7.91	&	7.53	&	7.59	&	$\dots$	&	$\dots$	&	$\dots$	\\
N$^{+}$/H$^+$	&	$\lambda$ 6584	&	5.82	&	$\dots$	&	$\dots$	&	$\dots$	&	$\dots$	&	5.74	&	6.46	&	6.51	&	6.68	&	6.48	&	6.55	&	8.57	&	9.41	&	8.25	\\
Ne$^{2+}$/H$^+$	&	$\lambda \lambda$ 3868, 3967	&	7.19	&	7.00	&	6.49	&	6.74	&	$\dots$	&	7.43	&	7.70	&	7.68	&	7.78	&	7.53	&	7.44	&	10.58	&	10.93	&	10.16	\\
Ne$^{3+}$/H$^+$	&	$\lambda \lambda$ 4715, 4726	&	$\dots$	&	$\dots$	&	$\dots$	&	$\dots$	&	$\dots$	&	8.68	&	8.84	&	$\dots$	&	$\dots$	&	$\dots$	&	$\dots$	&	12.02	&	12.59	&	11.58	\\
Ne$^{4+}$/H$^+$	&	$\lambda$ 3426	&	$\dots$	&	$\dots$	&	$\dots$	&	$\dots$	&	$\dots$	&	$\dots$	&	$\dots$	&	$\dots$	&	$\dots$	&	$\dots$	&	$\dots$	&	11.40	&	11.71	&	11.16	\\
Ar$^{3+}$/H$^+$	&	$\lambda \lambda$ 4711, 4740	&	$\dots$	&	$\dots$	&	$\dots$	&	$\dots$	&	$\dots$	&	5.91	&	6.22	&	$\dots$	&	$\dots$	&	$\dots$	&	$\dots$	&	$\dots$	&	9.00	&	8.18	\\
S$^{+}$/H$^+$	&	$\lambda \lambda$ 6716, 6730	&	5.24	&	$\dots$	&	$\dots$	&	$\dots$	&	$\dots$	&	4.43	&	4.53	&	5.14	&	$\dots$	&	5.32	&	$\dots$	&	$\dots$	&	7.23	&	$\dots$	\\
\hline
\end{tabular}
\end{center}
\end{table*}

\begin{table*}
\footnotesize
\begin{center}
\caption{Ionic abundances by number (12 + log\,$\frac{N(\mathrm{X^{+i})}}{N(\mathrm{H})}$) obtained from ORLs.}
\setlength{\tabcolsep}{0.5\tabcolsep}
\label{tbl:abn_ORLs}
%\begin{tabular}{l|c|ccccc|cccc|cc|ccc}
\begin{tabular}{lccccccccccccccc}
\hline
    &   &  
\multicolumn{5}{c}{Old nebula}  &  
\multicolumn{4}{c}{Eye-like structure}  &  
\multicolumn{2}{c}{High-velocity filament}  & 
\multicolumn{3}{c}{Disk knots and tails} \\
\cmidrule(lr){3-7}\cmidrule(lr){8-11}\cmidrule(lr){12-13}\cmidrule(lr){14-16}

        &    & W03A2 & N14A2 & W00A55$^\prime$ & W03A7 & N14A6 & W00A2 & W00A4 & W03A4 & N14A4 & W03A3 & N14A3 & W00A3 & W03A5 & W03A6 \\
\hline

He$^{+}$/H$^+$	&	$\lambda$ 4471	&	$\dots$	&	$\dots$	&	$\dots$	&	$\dots$	&	$\dots$	&	$\dots$	&	$\dots$	&	$\dots$	&	$\dots$	&	$\dots$	&	$\dots$	&	14.16	&	14.47	&	$\dots$	\\
He$^{+}$/H$^+$	&	$\lambda$ 5876	&	$\dots$	&	$\dots$	&	$\dots$	&	$\dots$	&	$\dots$	&	$\dots$	&	$\dots$	&	$\dots$	&	$\dots$	&	$\dots$	&	$\dots$	&	14.01	&	14.28	&	13.54	\\
He$^{+}$/H$^+$	&	$\lambda$ 6678	&	$\leq$9.43~	&	$\dots$	&	$\dots$	&	$\dots$	&	$\dots$	&	10.61	&	10.80	&	11.01	&	$\dots$	&	$\dots$	&	$\dots$	&	13.66	&	$\dots$	&	13.34	\\
He$^{2+}$/H$^+$	&	$\lambda$ 4686	&	10.97	&	11.01	&	11.08	&	11.07	&	11.04	&	11.11	&	11.16	&	10.98	&	10.98	&	11.03	&	10.97	&	13.86	&	14.37	&	13.75	\\

\hline
\end{tabular}
\end{center}
\end{table*}

The transitions probabilities and collision strengths were taken from the references listed in Table~\ref{tbl:abn_references}. The resultant ionic abundances are shown in Tables~\ref{tbl:abn_CELs} and \ref{tbl:abn_ORLs}.

\subsubsection{Total abundances}\label{sec:total_abundances}

The total He abundances were obtained by adding the ionic abundances of He$^{++}$ derived from the He~{\sc ii} $\lambda$4686 line and the average of the abundances of He$^+$ derived from all available He~{\sc i} lines. 
We note that the non-detection of He~{\sc i} emission lines in a number of apertures did not make possible to derive He$^+$ abundances for those.

As for the other elements, namely O, N, Ne and S, the contributions of ions inaccessible from optical spectra to the total abundances rely on ionization corrections factors (ICFs). 
Here we use those presented by \citet{Inglada2014} and \citet{Kingsburg1994} (hereinafter DM14 and KP94, respectively), which are specifically computed for PNe. 

Both are based on similar criteria to assess the likely ionization fractions using parameters derived from He~{\sc i}, He~{\sc ii}, [O~{\sc ii}] and [O~{\sc iii}] emission line ratios. 
We note that the applicability of the ICFs of DM14 is more restrictive than that of KB94, depending on the number of available lines. 
However, their effects on the final abundance estimates remain consistent with each other. 
In particular, DM14 define the ionic fractions $\nu$ and $\omega$, 
\begin{equation}
\centering
    \omega = \mathrm{ \frac{[O \hspace{0.05cm}\textsc{iii}]}  {[O \hspace{0.05cm}\textsc{ii}] + [O \hspace{0.05cm}\textsc{iii}]}}, \hspace{1.5cm}  \nu = \mathrm{ \frac{He \hspace{0.05cm}\textsc{ii}}{He\hspace{0.05cm}\textsc{ii} + He \hspace{0.05cm}\textsc{i}}}
\label{eqn:ionization_icf}
\end{equation}
that are used to validate the ICF selection (see table 3 in DM14).

We present in Table~\ref{tbl:abundances_total} the total abundances derived for all apertures where H$\beta$ is detected. 
The bottom rows of this table also present the values of $\omega$ and $\nu$ obtained from the different extraction regions.
We note that $\nu$ depends on He~{\sc i}, which is not detected in all apertures, particularly in none of the  
apertures probing the outer nebula.  
The aperture W03A2 (Southwest of the outer nebula), with the highest signal-to-noise ratio, has been used to estimate an upper limit of the He~{\sc i} emission lines intensities and corresponding He$^+$ ionic abundances (Table~\ref{tbl:abn_ORLs}) to assess the effects on the ICFs for the outer nebula.  
The O, N, Ne, and S then computed are marked by a "$\star$" sign in Table~\ref{tbl:abundances_total}.

The {\it Spitzer} IR spectrum provided us with measurements of the O$^{+3}$ $\lambda$27.6~$\mu$m emission line in the eye-like, HVF, and old nebula, which have been used to determine the total abundance of oxygen by direct addition. 
We notice that the abundances obtained this way are $\approx$0.6--1 dex lower than those obtained by ICFs, likely due to the fact that He~{\sc i} is not detected in the outer regions, resulting in $\nu=1$, which leads to overestimating the contribution of high-excitation species. 
Therefore, in the subsequent total abundances calculation for N, Ne, and S using ICFs, direct addition oxygen abundances were adopted.
The Ne abundances can also be determined by direct addition because there are emission lines of the main ions in the optical spectra.
As for the innermost region, the {\it Spitzer} spectrum was unfortunately saturated and it was not possible to measure the intensity of the O$^{+3}$ $\lambda$27.6~$\mu$m line there. 
Since the parameters $\omega$ and $\nu$ in this region are in the confidence range for the calculation of abundances from the ICFs, the total oxygen abundances can be considered reliable.

Table~\ref{tbl:abundances_total} includes many caveats regarding the total abundances of the different apertures, but we note that the abundances derived for apertures probing the same structural component using either the direct addition or the same ICF prescription are fairly consistent among them.  
This has allowed us to set the representative abundances for each structure presented in Table~\ref{tbl:final_abundances_total} that will be used to clarify the final results and to make a fair comparison with those of similar objects.
In particular we show in this table the abundances of the knots J1, J3 \citep{Wesson2003} and J4 \citep{Simpson2022} of A30 \citep[see][for notation]{Jacoby1979} and the knot of A\,58 \citep{Wesson2008}, as well as the Sun abundances \citep{Asplund2009}.

\begin{table*}
\caption{Total abundances computed using the ICFs listed in \citet{Inglada2014} (DM14) and \citet{Kingsburg1994} (KB94), and direct addition from ionic abundances, when possible.
The values of the parameters $\nu$ and $\omega$ (see Eq.~\ref{eqn:ionization_icf}) defined by \citet{Inglada2014} are presented in the bottom. 
Abundances in units such that log\,$N$(H) = 12.0. \\
Notes: \\
$^\mathrm{a}$: Since no He~{\sc i} emission lines are detected, the oxygen ICF in Equation A10 of KB94 is unity. \\
$^\mathrm{b}$: It is not recommended using the ICF of equation 10 in DM14 to compute the O abundance if $\nu$ is higher than 0.95. \\
$^\mathrm{c}$: It is not recommended using the ICF of equation 14 in DM14 to compute the N abundance if $\omega$ is higher than 0.95.\\
$^\mathrm{d}$: The ICF in equation 20 of DM14 was used instead of equation 17 in DM14, because the former takes Ne$^{4+}$ in addition to Ne$^{2+}$ to compute the Ne abundance.\\
%$^\mathrm{e}$: Only O$^{+}$ and O$^{2+}$ ionic abundances were available.\\
$^\star$Abundances were computed adopting an upper limit of the He$^+$ ionic abundance. 
}
\setlength{\tabcolsep}{0.5\tabcolsep}  
%\footnotesize
\begin{center}
\label{tbl:abundances_total}
\begin{tabular}{ll|ccccc|cccc|cc|ccc}
\hline
 &  &  \multicolumn{5}{c}{Old nebula}  &  \multicolumn{4}{c}{Eye-like structure}  &  \multicolumn{2}{c}{High-velocity filament}  & \multicolumn{3}{c}{Disk knots and tails} \\
\cmidrule(lr){3-7}\cmidrule(lr){8-11}\cmidrule(lr){12-13}\cmidrule(lr){14-16}

  & ICFs & W03A2 & N14A2 & W00A55$^\prime$ & W03A7 & N14A6 & W00A2 & W00A4 & W03A4 & N14A4 & W03A3 & N14A3 & W00A3 & W03A5 & W03A6 \\
\hline
\vspace{1mm} 
He/H  & Direct addition & $\geq$10.97 & $\geq$11.01 & $\geq$11.08 & $\geq$11.07 & $\geq$11.04 & 11.16 & 11.22 & 11.11 & $\geq$10.98 & $\geq$11.03 & $\geq$10.97 & 14.23 & 14.6 & 13.87\\
%        &  &  &  &  &   &  &  &  &  &  &  \\
%        &  &  &  &  &   &  &  &  &  &  &  \\

O/H	  & KB94 & 9.09$^\star$ & 7.78$^\mathrm{a}$ & 7.21$^\mathrm{a}$ & 7.22$^\mathrm{a}$ & 7.33$^\mathrm{a}$ & 8.69 & 8.83 & 8.56 & 8.27$^\mathrm{a}$ & 7.92$^\mathrm{a}$ & 8.0$^\mathrm{a}$ & 10.83 & 11.43 & 10.73 \\
      & DM14 & 8.92$^\mathrm{b}$$^\star$  & 9.01$^\mathrm{b}$ & 8.44$^\mathrm{b}$ & 8.45$^\mathrm{b}$ & 8.56$^\mathrm{b}$ & 8.82  & 8.94  & 8.61  & 9.5$^\mathrm{b}$  & 9.15$^\mathrm{b}$  & 9.23$^\mathrm{b}$  & 10.82  & 11.44  & 10.79 \\ \vspace{1mm} 
      %& Direct addition & 7.78 & 7.81 & 7.21$^\mathrm{e}$ & 7.22$^\mathrm{e}$ & 7.33$^\mathrm{e}$ & 8.12 & 8.28 & 8.21 & 8.43 & 8.07 & 8.15 & 10.67$^\mathrm{e}$ & 11.17$^\mathrm{e}$ & 10.33$^\mathrm{e}$ \\
      & Direct addition & 7.78 & 7.81 & \dots & \dots & \dots & 8.12 & 8.28 & 8.21 & 8.43 & 8.07 & 8.15 & \dots & \dots & \dots \\
      
N/H   & KB94 & 7.11$^\star$ & \dots & \dots & \dots & \dots & 7.68 & 7.78 & 7.60 & 7.80$^\mathrm{a}$ & 7.61$^\mathrm{a}$ & 7.45$^\mathrm{a}$ & 9.7 & 10.81 & 9.62  \\ \vspace{1mm} 
      & DM14 & 6.96$^\mathrm{c}$$^\star$  & \dots  & \dots & \dots & \dots & 7.53$^\mathrm{c}$  & 7.64$^\mathrm{c}$  & 7.48  & 7.66 & 7.47  & 7.31   & 9.61  & 10.71  & 9.56 \\

Ne/H  & KB94 & 7.23$^\star$ & 7.03$^\mathrm{a}$ & 6.49 & 6.74 & \dots & 7.54 & 7.77 & 7.76 & 7.98$^\mathrm{a}$ & 7.72$^\mathrm{a}$ & 7.67$^\mathrm{a}$ & 10.79 & 11.22 & 10.61 \\
      & DM14 & 7.59$^\star$ & 7.29 & 7.98 & 8.22 & \dots & 7.70 & 7.92 & 7.78 & 8.43 & 8.18 & 8.21 & 11.58$^\mathrm{d}$ & 11.95$^\mathrm{d}$ & 11.44$^\mathrm{d}$\\ \vspace{1mm} 
      & Direct addition & 7.19 & 7.0 & 6.49 & 6.74 & \dots & 8.71 & 8.87 & 7.68 & 7.78 & 7.53 & 7.44 & 12.12 & 12.65 & 11.73 \\

S/H	  & KB94 & 6.75$^\star$ & \dots& \dots& \dots & \dots & 6.39 & 6.05 & 6.49 & \dots & 6.66 & \dots & \dots & 8.70 & \dots \\ \vspace{1mm} 
      & DM14 & 6.32$^\star$ & \dots& \dots& \dots & \dots & 6.21 & 5.71 & 6.16 & \dots & 7.32 & \dots & \dots & 8.64 & \dots \\
\hline 
$\omega$ &	&	0.95 & 1.0& 1.0 & 1.0 & 1.0 & 0.99 & 0.95 & 0.91 & 0.89 & 0.9 & 0.82 & 0.89 & 0.93 & 0.89 \\
$\nu$    &  &	0.99 & 1.0& 1.0 & 1.0 & 1.0 & 0.91 & 0.87 & 0.74 & 1.0 & 1.0 & 1.0 & 0.42 & 0.59 & 0.75    \\
\hline
\end{tabular}
\end{center}
\end{table*}

\begin{table*}
\caption{
Total abundances and abundances ratios in different components of A\,78. 
For comparison, the abundances of clumps J1 and J3 \citep{Wesson2003} and J4 \citep{Simpson2022} of A\,30, those of the central knot of A\,58 \citep{Wesson2008}, the theoretical predictions of the abundances at the surface of the star after a VLTP event \citep{Miller2006}, and the Solar abundances \citep{Asplund2009} are included. 
Abundances are provided in units such that log\,$N$(H) = 12.0, whereas abundances ratios are by number. 
}
\setlength{\tabcolsep}{0.8\tabcolsep}  
%\footnotesize
\begin{center}
\label{tbl:final_abundances_total}
\begin{tabular}{l|cccc|ccc|c|c|c}
\hline
   &  \multicolumn{4}{c}{A78} & \multicolumn{3}{c}{A30} & A58 &  VLTP Event & Solar\\
\cmidrule(lr){2-5}\cmidrule(lr){6-8}\cmidrule(lr){9-9}\cmidrule(lr){10-10}\cmidrule(lr){11-11}

   & Old nebula (SW) &  Eye-like structure  &  High-velocity filament  & Disk knots and tails &  J1 & J3 & J4 & Central knot & Star Surface &\\
\hline
\vspace{1mm} 
He/H	& 11.03 & 11.16 & 11.00 & 14.23 & 13.03 & 13.07 & 12.62 & 12.51 & 14.21 & 10.93\\
\vspace{1mm} 
O/H		& 7.80 & 8.26 & 8.11 & 10.82 & 9.26 & 9.32 & 9.72 & 10.03 & 13.24 & 8.69 \\
\vspace{1mm}
N/H		& 7.03 & 7.62 & 7.46 & 9.85 & 8.88 & 8.90 &	9.18 & 9.21 & 12.67 & 7.83\\
\vspace{1mm} 
Ne/H	& 7.41 & 7.74 & 7.94 & 12.16 & 9.70 & 9.78 & 9.25 & 9.92 & 12.46 & 7.93\\
\vspace{1mm} 
S/H	    & 6.53 & 6.16 & $\dots$ & 8.09 & \dots & \dots & 6.96 & \dots & \dots & 7.12	\\
\hline 
N/O     & 0.17 & 0.23 & 0.22 &  0.10 & 0.41 & 0.38 & 0.28 &	0.15 & 0.26	& 0.13\\

Ne/O    & 0.41 & 0.3 & 0.67 & 21.87 &  2.75 & 2.88 &	0.33 & 0.77	& 0.16 & 0.17\\

S/Ne	& 0.13 & 0.026 & $\dots$  & 0.0003 & \dots & \dots & 0.005 & \dots & \dots & 0.15	\\
\hline
\end{tabular}
\end{center}
\end{table*}

\section{Discussion}
\label{sec:diss}

\subsection{Spatially varying extinction and dust distribution in A\,78}

The logarithmic extinction coefficient $c$(H$\beta$) has been found to vary across the different regions of A\,78, with values 
$\approx$1.5 for the inner VLTP ejecta, 
$\simeq$0.15 for the eye-like structure and Northwest region of the old nebula, 
$\simeq$0.38 for the HVF and 
% $\simeq$0.38 for the 
Southeast region of the old nebula, 
% $\simeq$0.15 for the 
and 0.03 for the East and West regions of the old nebula.

\citet{Cohen1977} presented the first extinction measurements towards A\,78, with $E(B-V)$ = 0.13 (or $c$(H$\beta$) = 0.18), from an optical spectrum of its CSPN extracted from a $4^{\prime\prime}\times2.7^{\prime\prime}$ aperture. 

\citet{Kaler1981} obtained a similar value for $c$(H$\beta$) of 0.15 using photometric measurements, although the errors were so large that the measurement was deemed not reliable. 
\cite{Kaler&Feidelman1984} derived more precise values for  $c$(H$\beta$) of 0.164 from the UV bump at 2200 \AA\ and 0.177 from optical data. 
These authors noted for the first time the position of the internal disk in A\,78 by comparing the extinctions obtained with that previously calculated for A\,30.
Meanwhile, \cite{Manchado1988} obtained spatially-resolved optical spectra of A\,78, including the outermost regions.  
Although the H$\alpha$ to H$\beta$ ratio among different regions provided by these authors certainly implied extinction variations, they only reported an average value $E(B-V)$=0.153, i.e., $c$(H$\beta$) of 0.22. 
All these values are consistent with the extinction derived here for the eye-like structure, which is reasonable given that it is the brightest extended nebular structure and presumably most easily probed by early observations.

The extinction of the inner VLTP ejecta derived in this work is significantly larger than any value reported previously. 
A seemingly high extinction of $c$(H$\beta$)=1.23 has recently been estimated from the Balmer decrement for the J4 knot \citep[see][for notation]{Jacoby1979} of A\,30 \citep{Simpson2022}, while that of the J1 and J3 knots have been derived to be 1.02 and 0.64, respectively, adopting temperatures from He line ratios \citep{Wesson2003}. 
The higher extinction of the inner ejecta of these born-again PNe can be expected given the significant presence of dust in their central region \citep[e.g.,][]{Cohen1974,Cohen1977}, with the dust spatially correlated with the disk knots as recently described by \citet{Toala2021}.  
The infrared emission of the dust in A\,30 has been attributed to hot ($\lesssim$160~K) and small carbon-rich dust spatially correlated with the inner disrupted disk, and larger and cooler ($<$80~K) dust located farther away up to the swept-up clumps and filaments \citep{Toala2021}.
The similar properties of the dust emission of A\,30 and A\,78 makes reasonable assuming they also have similar dust spatial distributions. 
Interestingly the dust coexists with X-ray-emitting material \citep{Toala2015}.

Finally, the old outer nebula in A\,78 shows remarkable extinction variations in correspondence with morphological and excitation variations. 
The lowest extinction values, $c$(H$\beta$)=0.03, are found towards the Northeast and Southwest regions registered by the W00A55$'$ aperture.  
Then the extinction increases towards the Northwest region to $c$(H$\beta$)=0.15 in the W03A7 and N14A6 apertures and very notably towards the Southeast region and at the HVF with $c$(H$\beta$)=0.38.  
There is a notable correlation between extinction and intensity of the [O~{\sc iii}] emission lines.  
If the lowest extinction value were assumed to correspond to the interstellar extinction towards A\,78, then the higher extinction at other regions of the outer nebula would require the presence of dust there.  
It must be noted, however, that there is no near-IR nor mid-IR dust emission beyond the spatial extent of the eye-like structure \citep{Toala2021}, although the broken morphology of the eye-like feature along the Southeast-Northwest direction suggests that material from the innermost region may leak up to the outer shell.  
We noted above that the dust model of A\,30, which implies that large dust grains with emission peaking at longer wavelengths are located further away from the CSPN \citep{Toala2021}, may also apply to A\,78. This hypothesis will be assess by future photoinization modeling of A\,78 including both gas and dust (Rodr\'\i guez-Gonz\'{a}lez et al., in preparation).

\subsection{Spatially varying physical properties and chemical abundances in A\,78}

An inspection of the physical properties and chemical abundances listed in Tables~\ref{tbl:lines_ratios_1}, \ref{tbl:abn_CELs}, \ref{tbl:abn_ORLs} and  \ref{tbl:final_abundances_total} immediately reveals notable variations in the excitation, physical properties and chemical abundances among the different structural components of A\,78.  
These are described below for the old H-rich nebula, the innermost and most recent VLTP ejecta, and the eye-like structure and other features indicative of mixing processes between H-poor and H-rich material. 

\subsubsection{Old H-rich nebula}

The line intensities derived from the spectra of the old hydrogen-rich nebula, presented in Table \ref{tbl:lines_ratios_1}, do not show significant variations in the $\approx$14 years elapsed between the different observations. 
The prevalence of [O~{\sc iii}] over [O~{\sc ii}], which is only detected in the W03A2 aperture, He~{\sc ii} over He~{\sc i}, and the He~{\sc ii}/H$\beta$ ratio of \textasciitilde 1 evidence a high-excitation nebula. 
Thus, if the nebula experienced a period of recombination after the born-again event, as is currently the case of the outer nebulae of Sakurai's Object and HuBi\,1 \citep{Guerrero_etal2018,Reichel_etal2022}, it has by now recovered due to the action of its CSPN strong ionizing flux ($T_{\mathrm{eff}}\approx$113~kK).

The average He/H abundances ($\simeq$0.11) are similar throughout the old nebula. 
As for the other elements, the abundances derived from spectra of apertures that allow their calculation are mostly consistent.  
Accordingly the abundances of the Southwest W03A2 aperture, with the highest signal to noise ratio, were chosen to be representative of the outer nebula in Table.~\ref{tbl:final_abundances_total}. 
 
The N/O ratio $\sim$0.17 is the average value for a type II PN \citep{Peimbert1978}. 
The S/Ne ratio also reveals a lack of S, which is otherwise typical among PNe \citep{Henry2004}.
The chemical enrichment models presented by \citet{Karakas2016} indicate that the progenitor star initial mass can be fairly restricted to be $\leq$2.5~M$_{\odot}$ by the He/H abundance, with the N/O ratio suggesting even a lower mass ($\lesssim$1~M$_{\odot}$) progenitor. 

\subsubsection{The VLTP ejecta}

The cometary knots of A\,78 distributed along the innermost disrupted disk and polar ejecta have been attributed to a VLTP event experienced by its CSPN.  
These features have experienced complex dynamical processes \citep[][]{Janis2022} with proper motions inconsistent with homologous expansion \citep{Fang2014}.  
As a result, the age derived from their angular expansion, which is distance independent, varies for different knots and filaments.  
Only a crude estimate $\approx$1000 yr can be obtained for the age of the VLTP event.

The most prominent property of the cometary knots of A\,78 is their extremely low H content, which is confirmed by the weak H$\alpha$ and H$\beta$ emission lines (Table~\ref{tbl:lines_ratios_1}) once that the overwhelmingly large contribution ($\simeq$90\%) of the He~{\sc ii} Pickering lines has been taken into account.  
We emphasise that the correction of the contribution of the He~{\sc ii} Pickering lines to the H~{\sc i} Balmer lines was accomplished here through an iterative process that accounted consistently for the dependence of the H~{\sc i} Balmer and He~{\sc ii} Pickering theoretical ratios on the temperature and density, which are coupled with the extinction.

The chemical abundances of the disk knots confirm their hydrogen deficiency, with abundances ratios by number of He/H$\sim$170, O/H$\sim$0.067 and N/H$\sim$0.007, i.e., around 1000 times larger than the abundances of the old H-rich nebula. 
We also note an excess of the Ne/H abundance, $\sim$1.4, and Ne to O abundance ratio by number of $\sim$22, which is otherwise seen in other born-again PNe, although to a lower extent.

This high Ne/O ratio is in stark contrast with the predictions of the surface of the star after undergoing a VLTP in a born-again event \citep[Ne/O$\sim$0.16,][]{Miller2006}. 

On the other hand, nova explosions result in Ne/O ratios of 1 or higher, which historically has led to a debate between the VLTP or nova event as the origin of born-again PNe.
Particularly for A58, \citet{Lau2011} proposed two binary scenarios, both with an oxygen-neon-magnesium (ONeMg) star as primary. 
In the first one, a nova event occurred shortly after a VLTP, while in the second a double common-envelope process is suggested. 
None of them fit perfectly with the observations, however, as they can only explain to some extent the O, C, and Ne rich ejecta. 
More recently, \citet{Janis2022} proposed that A\,30, a twin of A\,78, would have acquired its peculiar internal morphology after a common envelope process following the VLTP event.
The single or double common-envelope scenario would be a potential explanation for the morphology of A\,78 and to some extent for its chemical composition, but its extremely high Ne/O ratio indicates that there must be some additional process besides the nova, VLTP or common-envolope scenarios, implying either depletion of oxygen or an additional process for neon enrichment.

The abundances obtained for the H-poor knots of A\,78 are significantly higher, 1--2 dex, than those of the H-poor knots J1 and J3 \citep{Wesson2003} and J4 \citep{Simpson2022} of A\,30, and the central knot of A\,58 \citep{Wesson2008}.
This difference may be ascribed to the careful removal of the He~{\sc ii} contamination to the Balmer lines, whose fluxes has to be reckoned are subject to large uncertainties. 

Hydrogen determination aside, the trend followed by the abundances in A\,78 is N$<$O$<$Ne, alike the polar knots J1 and J3 of A\,30, but different from the trend N$<$Ne$<$O presented by its equatorial J4 knot. 
Such difference in the abundances of A\,30, if true, would suggest inhomogeneities in the eyected material in the VLTP.
This may be the case in A\,78 as well, but it has only been possible to calculate abundances of the knots located in the equatorial disk.
The N/O ratio, $\approx$0.1, similar to those obtained in the knots of A\,30 and A\,58, is expected in VLTP episodes \citep{Miller2006}. 
It is very notorious the extremely low S/Ne ratio $\approx$0.0003.  
This can be attributed to the Ne enrichment in this region, but also to the underestimation of the S abundances caused by the available ICFs in highly ionized nebular regions \citep{Henry2012}.  

We note that the temperature and density structure of the knots of A\,78 can be expected to be complex given the similarity with those in A\,30. 
These have been proposed to have dense, cold and chemically inhomogeneous cores in order to explain the temperature differences obtained from the ORL and CEL carbon lines \citep{Harrington1984}. 
The knots, with large CNO and He abundances, would be heated by the input photoionization but would immediately radiate away the energy excess through infrared fine-structure lines.  
This would produce a stratification where the knot outermost regions are highly ionized, whereas its innermost regions are denser and colder. 
These regions with different physical conditions cannot be isolated by the spatial resolution of ground-based observations, but we note that low ionization species find in the innermost core of the knots a shelter from the harsh CSPN ionizing flux.  
The detection of He~{\sc i} emission lines in the innermost knots of A\,78 thus support the stratification of chemical abundances and physical conditions within the knots, as He would be otherwise doubly ionized by the strong ionizing flux of the CSPN as in the eye-like structure and outer nebula.

\subsubsection{Mixing regions}

All of the H-deficient structures inside the old H-rich nebula in A\,78 can be considered to be remnants of the VLTP event experienced by its CSPN about 1000 years ago \citep{Fang2014}, either directly from the VLTP event, such as the cometary knots, or as the result of the interaction between the H-poor ejecta and the H-rich old nebula, giving rise to the eye-like structure very alike the petal-like structure seen in A\,30. 

The expected gradual transition between the material of the VLTP and the material of the old nebula, hinted by \citet{Manchado1988}, is clearly confirmed here, with chemical abundances of O, N and Ne in the eye-like structure, 2--4 times larger than those in the old nebula.

The notorious high-velocity filament HVF, with a velocity clearly different from that of the eye-like structure \citep{Fang2014}, could be a fragment of pristine ejecta that has reached the outermost nebular regions. 
Although that might be the case, its physical and chemical properties do not reveal any appreciable difference with the eye-like structure.
Spectroscopic observations at higher spatial and/or spectral resolution would be required to avoid the emission of this interesting feature to be diluted by the background nebular emission that is also registered by the spectral aperture used here.

\section{Summary}
\label{sec:summary}

We present here the analysis of long-slit optical and Spitzer mid-infrared spectra of the born-again PN A78. The spectra from 3 different epochs have allowed us to determine the ionic and total abundances from the CELs for the main structures in this PN, namely, the old H-rich nebula, the eye-like structure and the central cometary knots formed as a result of a born-again event. 
Our main findings can be sumarized as follows:

\begin{enumerate}

\item 
The temperature of the nebula derived from [O~{\sc iii}] lines remains relatively uniform, around $T_\mathrm{e}=$18,000 K, with a possible increase towards the central regions ($T_\mathrm{e}\approx$23,000 K). 
The densities derived from the [O~{\sc ii}] $\lambda\lambda$3726,3729 doublet range from 100 (low density limit) to 1000 cm$^{-3}$. 
The temperature and density structure of these knots is complex, as those in A\,30, with presumably high density and cold cores at their heads as revealed by emission in low-ionization lines (e.g., He~{\sc i}).

\item 
We have found an inhomogeneous extinction over the nebula unnoticed in previous spectroscopic investigations. 
In the eye-like structure and Northwest corner of the old H-rich nebula, the extinctions are similar and agree with previous estimates, while in the Southeast regions of the old nebula the extinction is higher. 
This increase may be due to the diffusion of dust leaking from the eye-like structure to the outer regions. 
The extinction in the central knots, even subject to large uncertainties, is undoubtedly higher than in the rest of the nebula, consistent with the large infrared dust emission.

\item The chemical abundances of the old nebula set A\,78 as a type II nebula following Peimbert's classification. 
A comparison of these chemical abundances with models of chemical enrichment indicates that the mass of the progenitor star is lower than 2.5~M$_{\odot}$, probably even $\lesssim$1~M$_{\odot}$.

\item 
The chemical abundances in the central knots are extremely H deficiency, with a low N/O ratio, $\approx$0.10, and greatly enhanced Ne abundances and Ne/O ratio typical of other born-again PNe.  
These abundances are neither consistent with a VLTP event nor a nova explosion, and have been suggested to result from post-VLTP strong binary interactions \citep{Lau2011,Janis2022}.

\item 
The eye-like structure and the high-velocity filament towards the Southeast show evidence of mixed material between the H-poor ejecta and the H-rich surrounded nebula. 

\end{enumerate}

\section*{Acknowledgments} 

BMM and MAG are funded by grant PGC 2018-102184-B-I00 of the Ministerio de Educaci\'{o}n, Innovaci\'{o}n y Universidades cofunded with FEDER funds. 
JAT thanks Direcci\'{o}n General de Asuntos del Personal Acad\'{e}mico (DGAPA) of the Universidad Nacional Aut\'{o}noma de M\'{e}xico (UNAM, Mexico) project IA101622
and the Visiting-Incoming programme of the IAA-CSIC supported by the State Agency for Research of the Spanish MCIU through the Center of Excellence Severo Ochoa award to the Instituto de Astrof\'{i}sica de Andaluc\'{i}a (SEV-2017-0709)through the Centro de Excelencia Severo Ochoa (Spain). 
This work has made extensive use of NASA's Astrophysics Data System (ADS). 

This paper is partially based on ground-based observations from 
(1) ALFOSC at the NOT and (2) ISIS at the WHT of the Observatorio de El Roque de los Muchachos (ORM).
This research is also based on observations made with the NASA/ESA {\it HST} obtained from the Space Telescope Science Institute, which is operated by the Association of Universities for Research in Astronomy, Inc., under NASA contract NAS 5–26555. 
This woek make use of Spitzer IR observations, which was operated by the Jet Propulsion Laboratory, California Institute of Technology, under a contract with NASA.

\section*{Data availability}

The processed data presented in this article will be shared on reasonable request to the corresponding author.

%%%%%%%%%%%%%%%%%%%%%%%%%%%%%%%%%%%%%%%%%%%%%%%%%%



%\bibliographystyle{mnras}

\begin{thebibliography}{99}

\bibitem[Asplund et al.(2009)]{Asplund2009} 
Asplund, M., Grevesse, N., Sauval, A.~J., et al.\ 2009, \araa, 47, 481. doi:10.1146/annurev.astro.46.060407.145222

\bibitem[Borkowski et al.(1993)]{Borkowski1993} Borkowski, K.~J., Harrington, J.~P., Tsvetanov, Z., et al.\ 1993, \apjl, 415, L47

\bibitem[Clegg et al.(1993)]{Clegg1993} Clegg, R.~E.~S., Devaney, M.~N., Doel, A.~P., et al.\ 1993, Planetary Nebulae, 155, 388

\bibitem[Cohen \& Barlow(1974)]{Cohen1974} Cohen, M. \& Barlow, M.~J.\ 1974, \apj, 193, 401

\bibitem[Cohen et al.(1977)]{Cohen1977} Cohen, M., Hudson, H.~S., O'Dell, S.~L., et al.\ 1977, \mnras, 181, 233

\bibitem[Dance et al.(2013)]{Dance2013} Dance, M., Palay, E., Nahar, S.~N., et al.\ 2013, \mnras, 435, 1576. doi:10.1093/mnras/stt1398

\bibitem[Delgado-Inglada et al.(2014)]{Inglada2014} Delgado-Inglada, G., Morisset, C., \& Stasi{\'n}ska, G.\ 2014, \mnras, 440, 536. 

\bibitem[Fang et al.(2014)]{Fang2014} Fang, X., Guerrero, M.~A., Marquez-Lugo, R.~A., et al.\ 2014, \apj, 797, 100

\bibitem[Froese Fischer \& Tachiev(2004)]{FroeseFischer2004} Froese Fischer, C. \& Tachiev, G.\ 2004, Atomic Data and Nuclear Data Tables, 87, 1. doi:10.1016/j.adt.2004.02.001

\bibitem[Galavis et al.(1997)]{Galavis1997} Galavis, M.~E., Mendoza, C., \& Zeippen, C.~J.\ 1997, \aaps, 123, 159. doi:10.1051/aas:1997344

\bibitem[Giles(1981)]{Giles1981} Giles, K.\ 1981, \mnras, 195, 63P. doi:10.1093/mnras/195.1.63P

\bibitem[Godefroid \& Fischer(1984)]{Godefroid1984} Godefroid, M. \& Fischer, C.~F.\ 1984, Journal of Physics B Atomic Molecular Physics, 17, 681. doi:10.1088/0022-3700/17/5/008

\bibitem[\protect\citeauthoryear{Guerrero et al.}{2018}]{Guerrero_etal2018} 
Guerrero M.~A., Fang X., Miller Bertolami M.~M., Ramos-Larios G., Todt H., Alarie A., Sabin L., et al., 2018, NatAs, 2, 784
% doi:10.1038/s41550-018-0551-8

\bibitem[Gvaramadze et al.(2020)]{Gvaramadze2020} 
Gvaramadze, V.~V., Kniazev, A.~Y., Gr{\"a}fener, G., et al.\ 2020, \mnras, 492, 3316

\bibitem[Harrington \& Feibelman(1984)]{Harrington1984} 
Harrington, J.~P. \& Feibelman, W.~A.\ 1984, \apj, 277, 716. doi:10.1086/161743

\bibitem[Harrington et al.(1995)]{Harrington1995} 
Harrington, J.~P., Borkowski, K.~J., \& Tsvetanov, Z.\ 1995, \apj, 439, 264. 
% doi:10.1086/175169

\bibitem[Heap(1979)]{Heap1979} 
Heap, S.~R.\ 1979, Mass Loss and Evolution of O-Type Stars, 83, 99

\bibitem[Henry et al.(2004)]{Henry2004} 
Henry, R.~B.~C., Kwitter, K.~B., \& Balick, B.\ 2004, \aj, 127, 2284. 
% doi:10.1086/382242

\bibitem[Henry et al.(2012)]{Henry2012} 
Henry, R.~B.~C., Speck, A., Karakas, A.~I., et al.\ 2012, \apj, 749, 61. 
% doi:10.1088/0004-637X/749/1/61


\bibitem[\protect\citeauthoryear{Herald \& Bianchi}{2004}]{HB2004} 
Herald J.~E., Bianchi L., 2004, ApJ, 609, 378. 
% doi:10.1086/421010

\bibitem[Iben et al.(1983)]{Iben1983} 
Iben, I., Kaler, J.~B., Truran, J.~W., et al.\ 1983, \apj, 264, 605. 
% doi:10.1086/160631

\bibitem[Jacoby(1979)]{Jacoby1979} 
Jacoby, G.~H.\ 1979, \pasp, 91, 754

\bibitem[Jacoby \& Ford(1983)]{Jacoby1983} 
Jacoby, G.~H. \& Ford, H.~C.\ 1983, \apj, 266, 298. 
% doi:10.1086/160779

\bibitem[Kaler(1981)]{Kaler1981} 
Kaler, J.~B.\ 1981, \apjl, 250, L31

\bibitem[Kaler \& Feibelman(1984)]{Kaler&Feidelman1984} 
Kaler, J.~B. \& Feibelman, W.~A.\ 1984, \apj, 282, 719

\bibitem[Karakas \& Lugaro(2016)]{Karakas2016} 
Karakas, A.~I. \& Lugaro, M.\ 2016, \apj, 825, 26. 
% doi:10.3847/0004-637X/825/1/26

\bibitem[Kingsburgh \& Barlow(1994)]{Kingsburg1994} 
Kingsburgh, R.~L. \& Barlow, M.~J.\ 1994, \mnras, 271, 257. 
% doi:10.1093/mnras/271.2.257

\bibitem[Kisielius et al.(2009)]{Kisielius2009} Kisielius, R., Storey, P.~J., Ferland, G.~J., et al.\ 2009, \mnras, 397, 903. 
% doi:10.1111/j.1365-2966.2009.14989.x

\bibitem[Luridiana et al.(2015)]{pyneb2015} Luridiana, V., Morisset, C., \& Shaw, R.~A.\ 2015, \aap, 573, A42. 
% doi:10.1051/0004-6361/201323152

\bibitem[Lau et al.(2011)]{Lau2011} 
Lau, H.~H.~B., De Marco, O., \& Liu, X.-W.\ 2011, \mnras, 410, 1870. 
% doi:10.1111/j.1365-2966.2010.17568.x

\bibitem[Manchado et al.(1988)]{Manchado1988} 
Manchado, A., Pottasch, S.~R., \& Mampaso, A.\ 1988, \aap, 191, 128

\bibitem[McLaughlin \& Bell(2000)]{McLaughlin2000} McLaughlin, B.~M. \& Bell, K.~L.\ 2000, Journal of Physics B Atomic Molecular Physics, 33, 597. 
% doi:10.1088/0953-4075/33/4/301

\bibitem[Meaburn et al.(1998)]{Meaburn1998} 
Meaburn, J., Lopez, J.~A., Bryce, M., et al.\ 1998, \aap, 334, 670

\bibitem[Miller Bertolami et al.(2006)]{Miller2006} 
Miller Bertolami, M.~M., Althaus, L.~G., Serenelli, A.~M., et al.\ 2006, \aap, 449, 313. 
% doi:10.1051/0004-6361:20053804

\bibitem[Montoro-Molina et al.(2022a)]{MM2022a} 
Montoro-Molina, B., Guerrero, M.~A., P{\'e}rez-D{\'\i}az, B., et al.\ 2022, \mnras, 512, 4003. 
% doi:10.1093/mnras/stac336

\bibitem[\protect\citeauthoryear{Montoro-Molina et al.}{2022b}]{M-M_etal2022} 
Montoro-Molina B., Guerrero M.~A., Toal{\'a} J.~A., Rodr{\'\i}guez-Gonz{\'a}lez J.~B., 2022, ApJ, 934, 18. 
% doi:10.3847/1538-4357/ac771b

\bibitem[Ohnaka \& Jara Bravo(2022)]{Ohnaka2022} Ohnaka, K. \& Jara Bravo, B.~A.\ 2022, \aap, 668, A119

%\bibitem[Peimbert(1981)]{Peimbert1981} Peimbert, M.\ 1981, Physical Processes in Red Giants, 88, 409. doi:10.1007/978-94-009-8492-9\_45

\bibitem[\protect\citeauthoryear{Peimbert}{1978}]{Peimbert1978} 
Peimbert M., 1978, IAUS, 76, 215

\bibitem[\protect\citeauthoryear{Phillips, Cuesta, \& Kemp}{2005}]{PCK2005} 
Phillips J.~P., Cuesta L., Kemp S.~N., 2005, \mnras, 357, 548. 
% doi:10.1111/j.1365-2966.2005.08664.x

\bibitem[Ramsbottom et al.(1997)]{Ramsbottom1997} 
Ramsbottom, C.~A., Bell, K.~L., \& Keenan, F.~P.\ 1997, \mnras, 284, 754. 
% doi:10.1093/mnras/284.3.754

\bibitem[\protect\citeauthoryear{Reichel et al.}{2022}]{Reichel_etal2022} 
Reichel M., Kimeswenger S., van Hoof P.~A.~M., Zijlstra A.~A., Barr{\'\i}a D., Hajduk M., Van de Steene G.~C., et al., 2022, ApJ, 939, 103. doi:10.3847/1538-4357/ac90c4

\bibitem[Rodr{\'\i}guez-Gonz{\'a}lez et al.(2022)]{Janis2022} 
Rodr{\'\i}guez-Gonz{\'a}lez, J.~B., Santamar{\'\i}a, E., Toal{\'a}, J.~A., et al.\ 2022, \mnras, 514, 4794. 
% doi:10.1093/mnras/stac1697

\bibitem[Rynkun et al.(2019)]{Rynkun2019} Rynkun, P., Gaigalas, G., \& J{\"o}nsson, P.\ 2019, \aap, 623, A155. 
% doi:10.1051/0004-6361/201834931

\bibitem[Sch\"{o}nberner(1979)]{Schonberner1979} Sch\"{o}nberner, D.\ 1979, \aap, 79, 108

\bibitem[Simpson et al.(2022)]{Simpson2022} 
Simpson, J., Jones, D., Wesson, R., et al.\ 2022, Research Notes of the American Astronomical Society, 6, 4

\bibitem[Smith et al.(2007)]{Smith2007} Smith, J.~D.~T., Armus, L., Dale, D.~A., et al.\ 2007, \pasp, 119, 1133

\bibitem[Storey et al.(2014)]{Storey2014} Storey, P.~J., Sochi, T., \& Badnell, N.~R.\ 2014, \mnras, 441, 3028. 
% doi:10.1093/mnras/stu777

\bibitem[Tayal(2011)]{Tayal2011} Tayal, S.~S.\ 2011, \apjs, 195, 12. doi:10.1088/0067-0049/195/2/12

\bibitem[Tayal \& Zatsarinny(2010)]{Tayal2010} Tayal, S.~S. \& Zatsarinny, O.\ 2010, \apjs, 188, 32. doi:10.1088/0067-0049/188/1/32

\bibitem[Toal{\'a} et al.(2021)]{Toala2021} 
Toal{\'a}, J.~A., Jim{\'e}nez-Hern{\'a}ndez, P., Rodr{\'\i}guez-Gonz{\'a}lez, J.~B., et al.\ 2021, \mnras, 503, 1543. 
% doi:10.1093/mnras/stab593

\bibitem[Toal{\'a} et al.(2015)]{Toala2015} Toal{\'a}, J.~A., Guerrero, M.~A., Todt, H., et al.\ 2015, \apj, 799, 671411.3837

\bibitem[Tody(1993)]{Tody1993} Tody, D.\ 1993, Astronomical Data Analysis Software and Systems II, 52, 173

\bibitem[Wesson et al.(2003)]{Wesson2003} 
Wesson, R., Liu, X.-W., \& Barlow, M.~J.\ 2003, \mnras, 340, 253

\bibitem[Wesson et al.(2008)]{Wesson2008} 
Wesson, R., Barlow, M.~J., Liu, X.-W., et al.\ 2008, \mnras, 383, 1639. doi:10.1111/j.1365-2966.2007.12683.x

\bibitem[Zeippen(1982)]{Zeippen1982} Zeippen, C.~J.\ 1982, \mnras, 198, 111. doi:10.1093/mnras/198.1.111


\end{thebibliography}


\clearpage

%%%%%%%%%%%%%%%%%%%%%%%%%%%%%%%%%%%%%%%%%%%%%%%%%%

%%%%%%%%%%%%%%%%% APPENDICES %%%%%%%%%%%%%%%%%%%%%

\appendix
\section{Surface brightness profiles}
\label{apex:A}

\begin{figure*}
\centering
\includegraphics[width=1\linewidth]{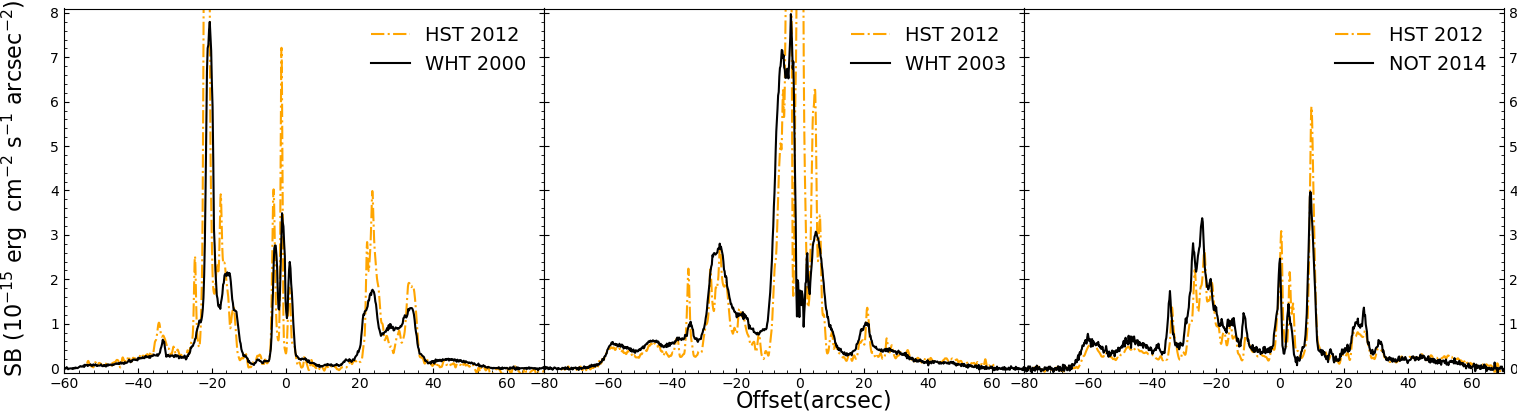} 
\caption{
Comparison of surface brightness profiles of the [O~{\sc iii}] $\lambda$5007 \AA\ emission line extracted from the spectroscopic WHT ISIS 2000.50 and 2003.58, and NOT ALFOSC 2014.55 observations (solid black lines) and HST WFC3 F502N (2012.89.44) image (dash-dotted orange lines).
}
\label{fig:A78_profiles}
\end{figure*}

Continuum-subtracted surface brightness profiles of [O~{\sc iii}] $\lambda$5007 extracted from the 2000.50 and 2003.58 WHT and 2014.55 NOT spectra (solid black lines in Figure~\ref{fig:A78_profiles}). 
The continuum-subtraction, effectively removing the contribution of stars in the slit, was estimated from both sides of the emission line. 
In the case of 2000.50 WHT profile, the aperture did not pass through the CSPN and its location in A\,78 relies on nebular features such as the central knots or the eye-like structure.

These surface brightness profiles are compared with those extracted from the HST WFC3 F502N image along pseudo-slits with the same widths and position angles of the spectroscopic observations (orange dashed-dotted lines in Fig.~\ref{fig:A78_profiles}) to assess possible time variation and to determine the exact location of each slit onto the nebula. 
The HST profiles have been smoothed to make a fair comparison, however, the emissions from both the central star and the background stars is not subtracted.

\section{One-dimensional Spectra}
\label{apex:B}

The spectra extracted from the apertures shown in Figure~\ref{fig:A78_slit} are presented in Figures~\ref{fig:A78_spec_w0}, \ref{fig:A78_spec_w3}, and \ref{fig:A78_spec_n14}, corresponding to the WHT 2000.50 and 2003.58, and NOT 2014.55 observations, respectively. 
Sky lines are marked by red ticks in those spectra with an imperfect sky background subtraction. 
The two-dimensional spectra obtained in the WHT present a diffraction pattern that was not possible to avoid and the subsequent one-dimensional spectra present this effect as a sinusoidal continuum. 
This is specially noticeable in spectra of the faintest regions. 
Despite the fact that the apertures selected for the extraction of the spectra of the knots are separated from the CSPN (Fig.~\ref{fig:A78_slit}), both the seeing and the spatial resolution of the observations caused the emission of the star to smear within the nebular spectra. 
The stellar features in those spectra have not been used during the data analysis.

\begin{figure*}
\centering
\includegraphics[width=0.95\linewidth]{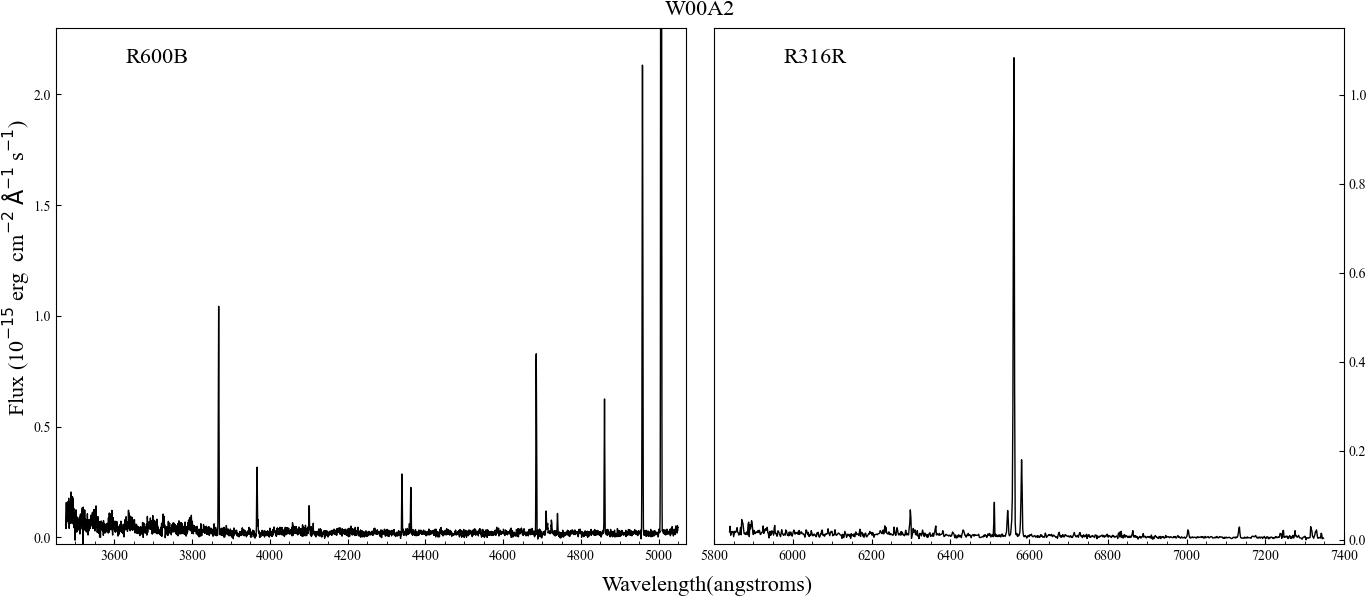}
\\
\includegraphics[width=0.95\linewidth]{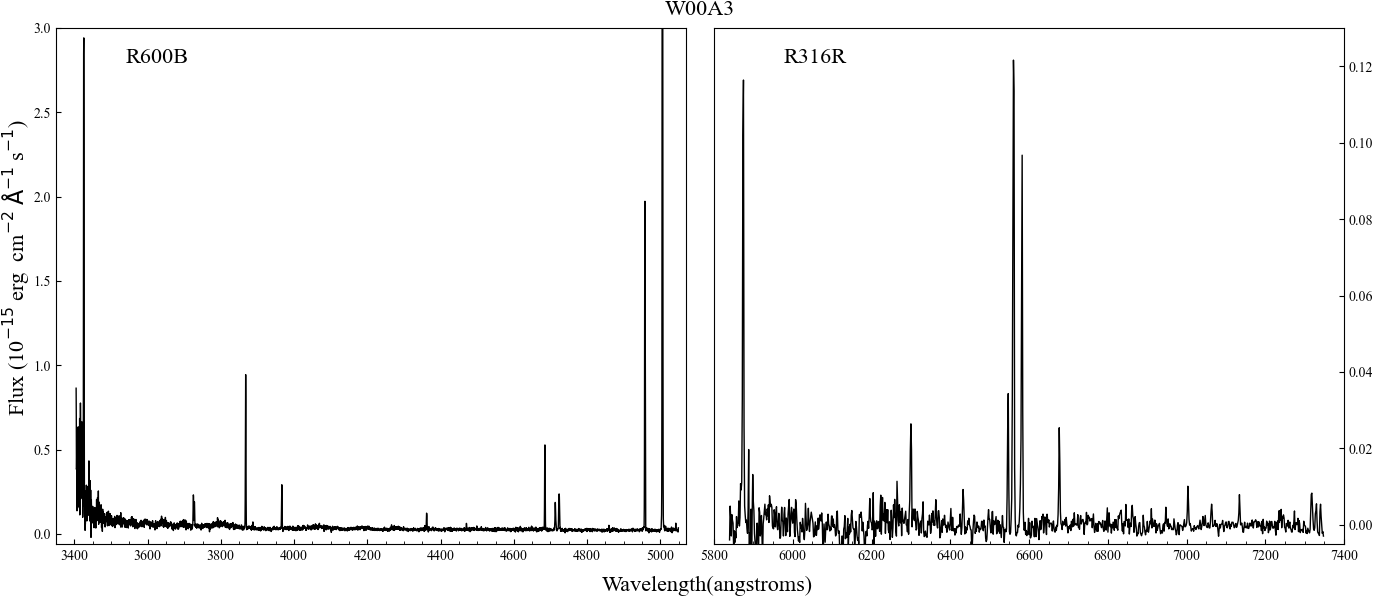}
\caption{
Spectra of the apertures extracted from the 2000.50 observations at WHT. 
}
\label{fig:A78_spec_w0}
\end{figure*}

\begin{figure*}
\centering
\ContinuedFloat
\includegraphics[width=0.95\linewidth]{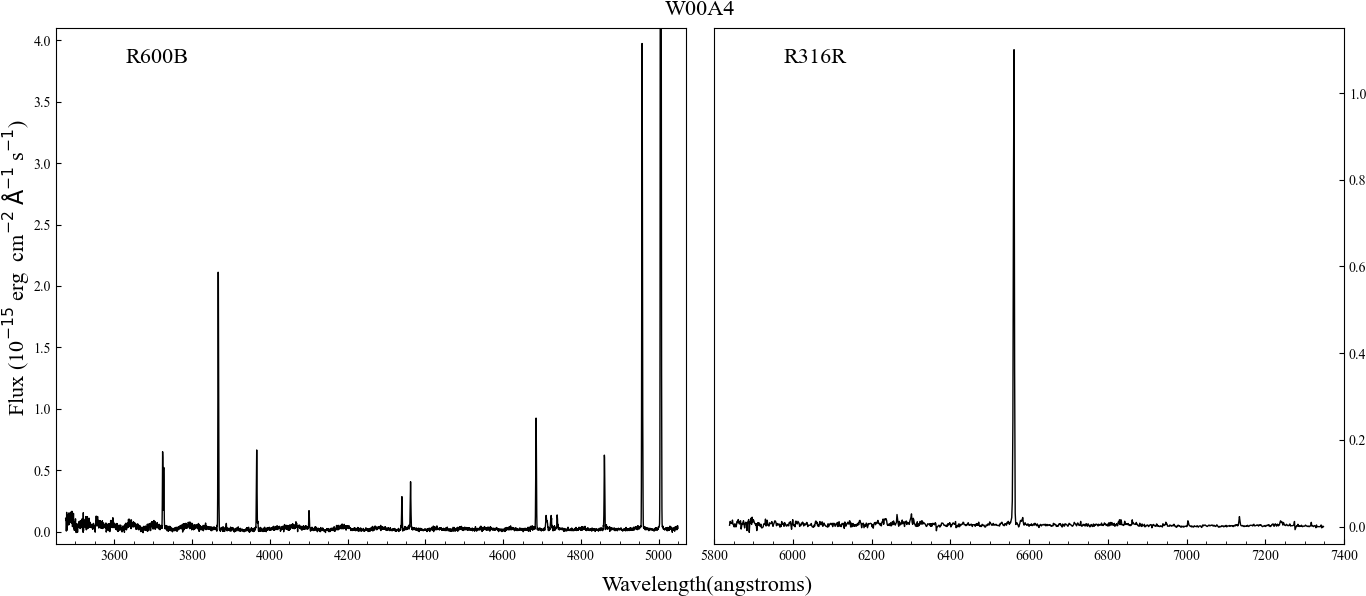}
\\
\includegraphics[width=0.95\linewidth]{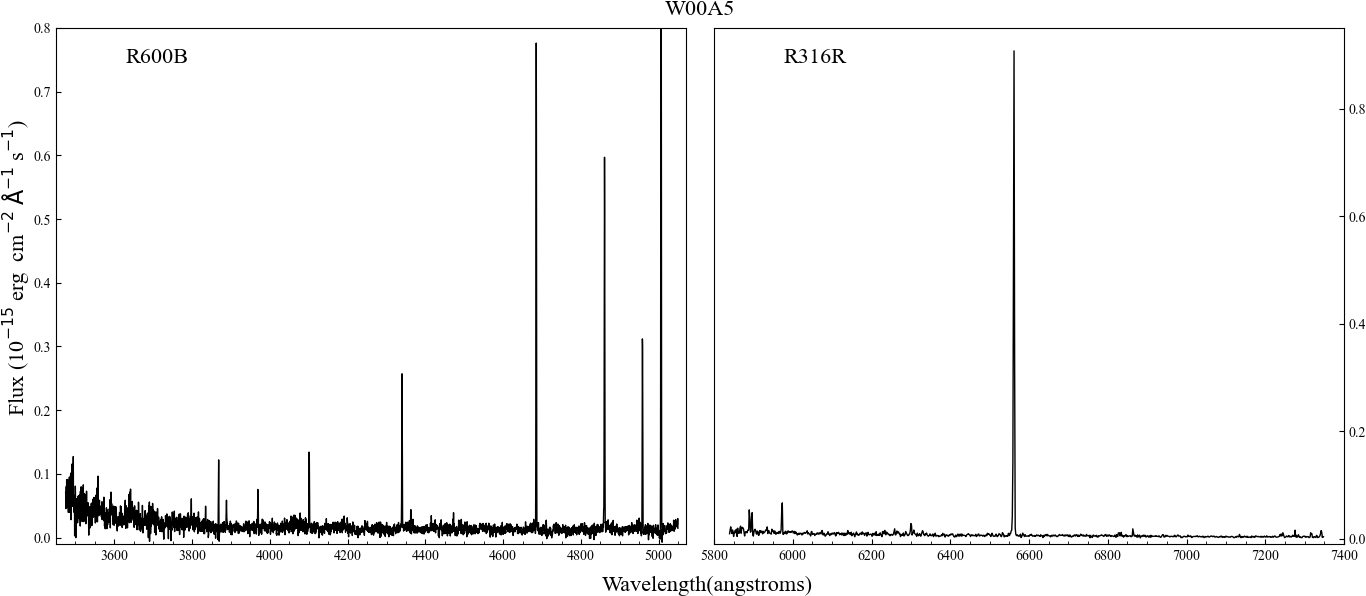}
\caption{
Continued. 
% Spectra of the apertures extracted from the 2000.50 observations at WHT. 
}
\end{figure*}

\begin{figure*}
\centering
\includegraphics[width=0.95\linewidth]{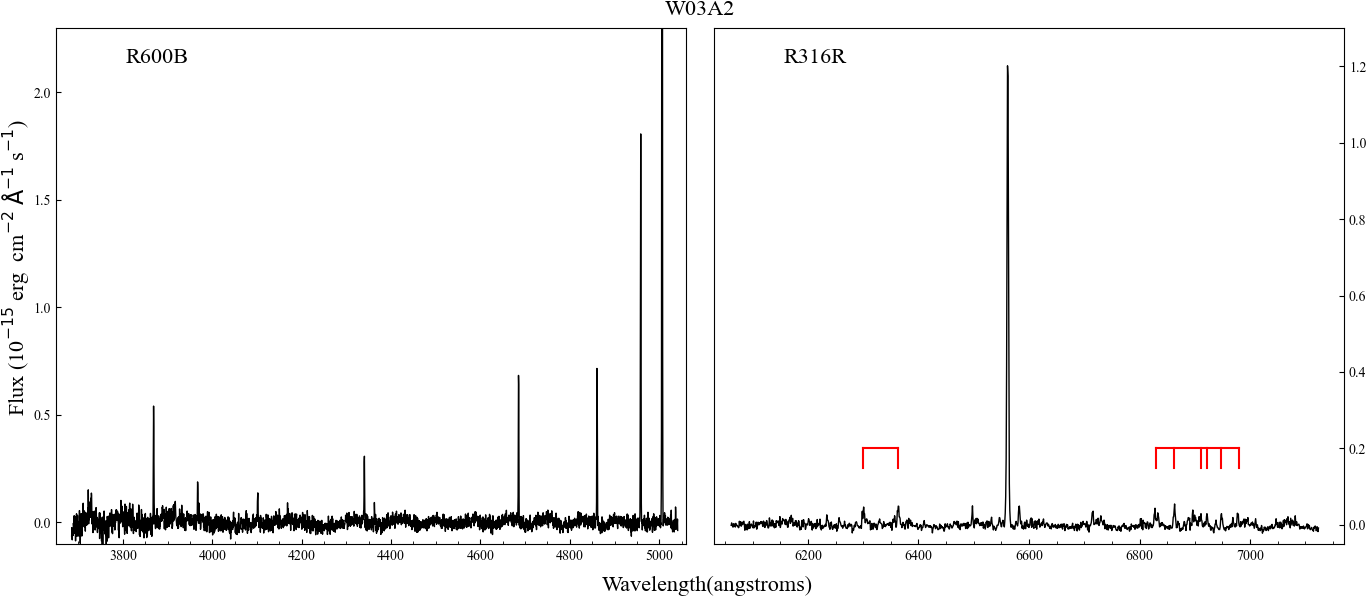} 
\\
\includegraphics[width=0.95\linewidth]{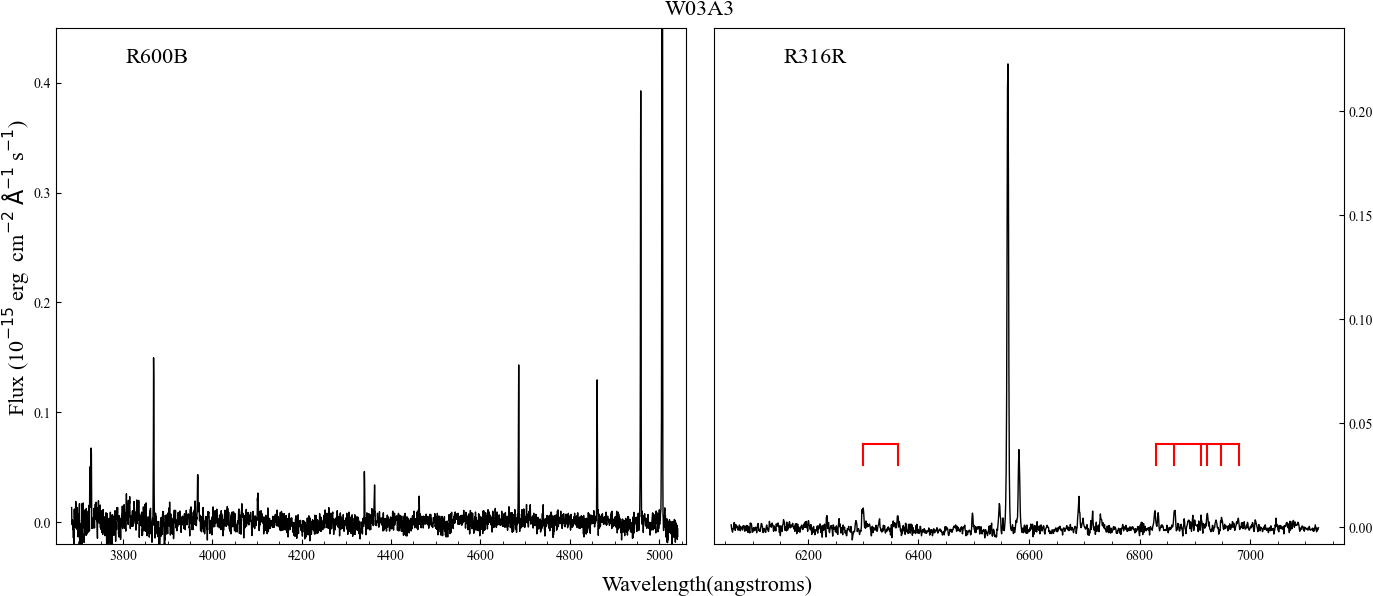}
\\
\includegraphics[width=0.95\linewidth]{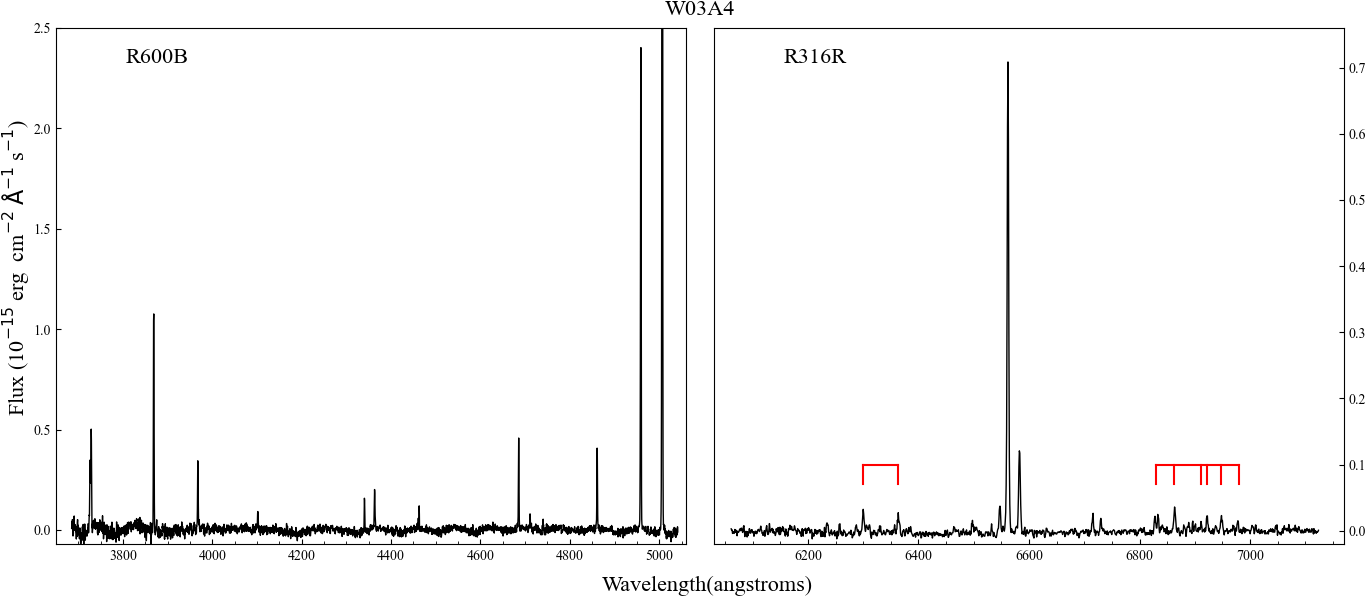}
\caption{
Spectra of the apertures extracted from the 2003.58 observations at WHT. 
The sky emission lines are marked by red ticks to avoid confusing them with nebular emission lines. 
}
\label{fig:A78_spec_w3}
\end{figure*}

\begin{figure*}
\centering
\ContinuedFloat
\includegraphics[width=0.95\linewidth]{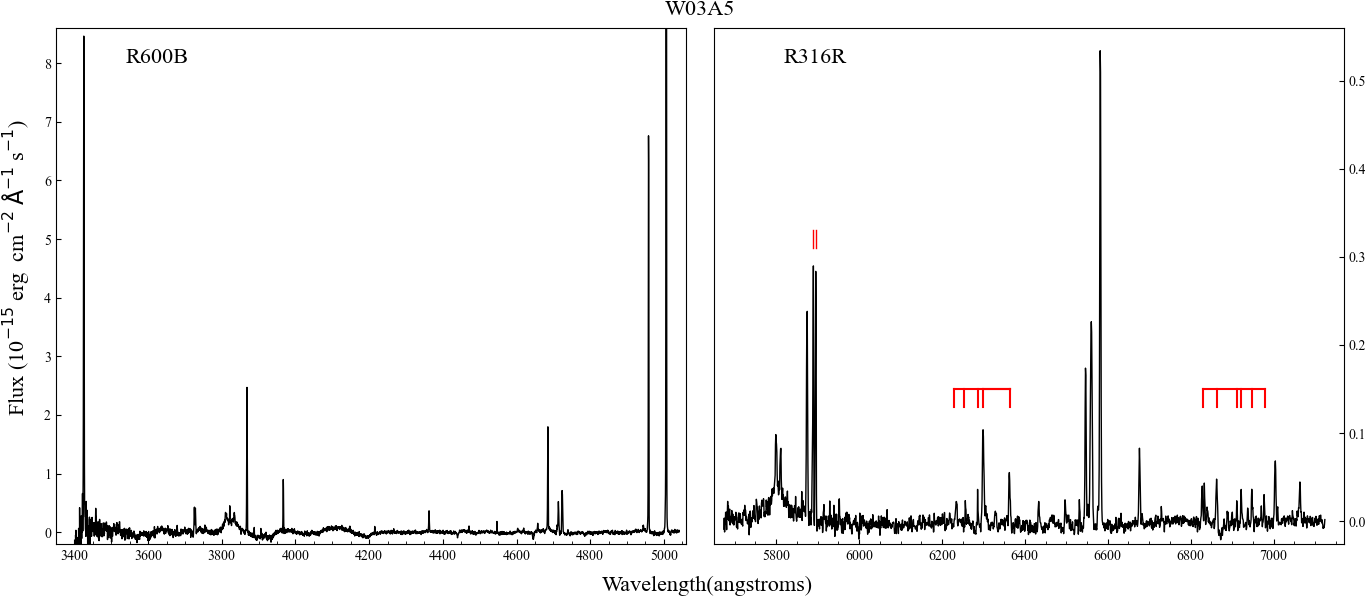}
\\
\includegraphics[width=0.95\linewidth]{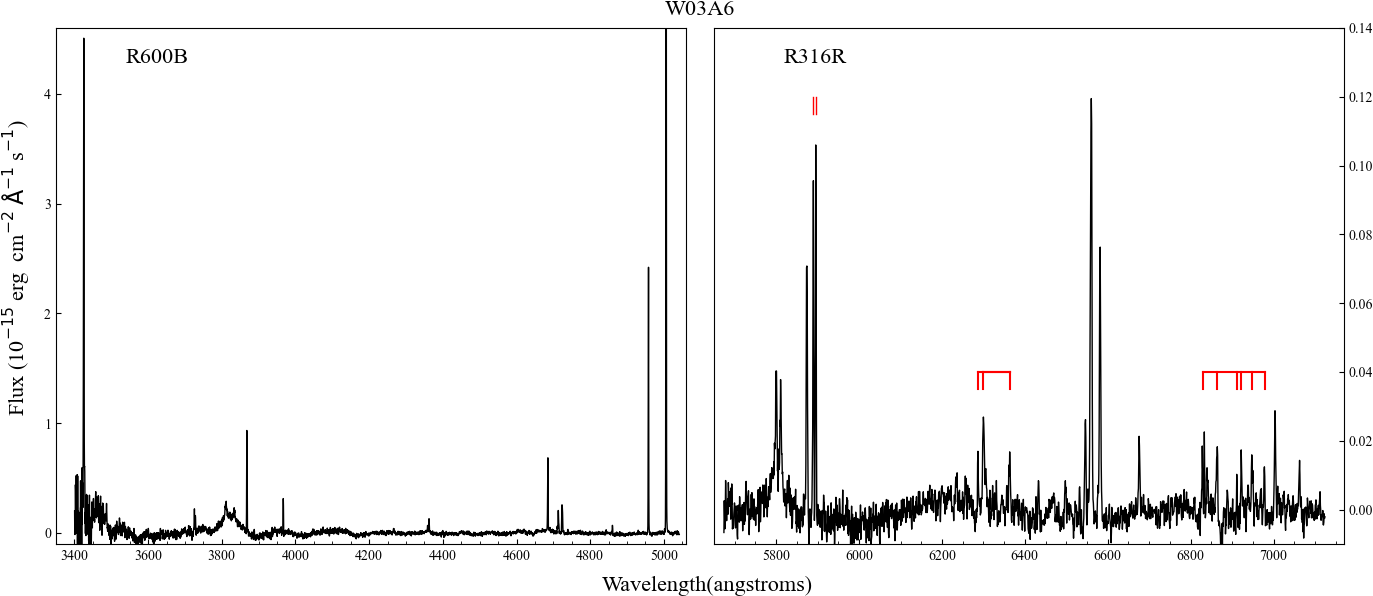}
\\
\includegraphics[width=0.95\linewidth]{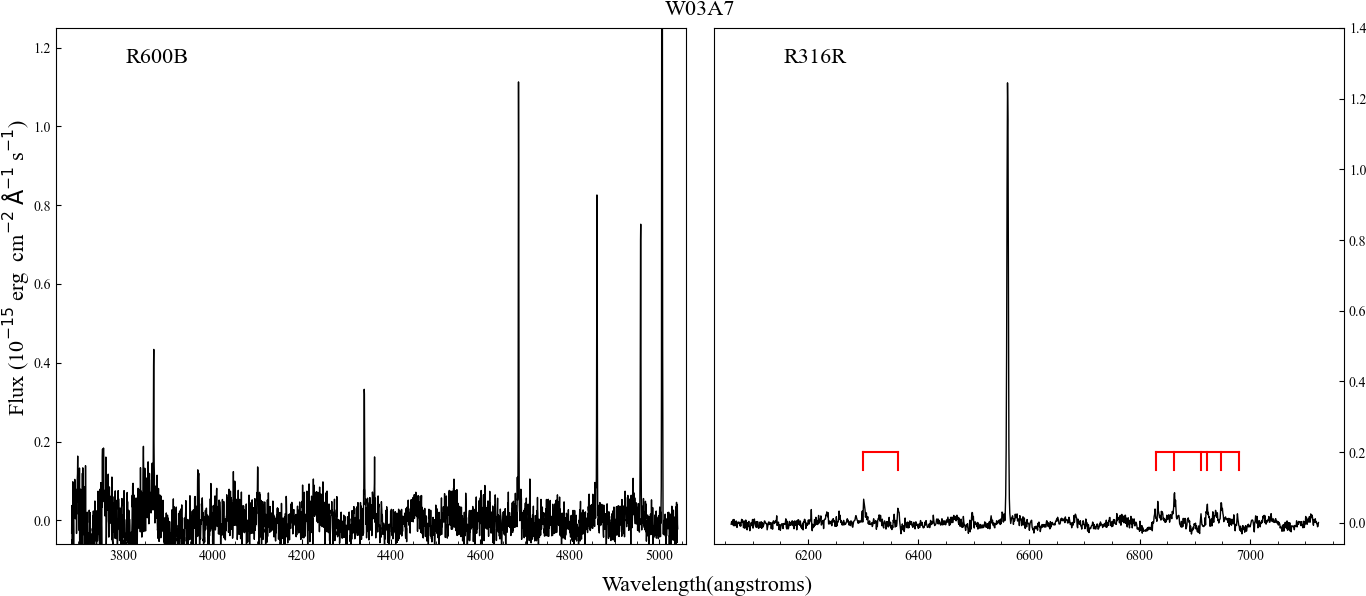}
\caption{
Continued.}
\end{figure*}

%%%%%%%%%%%%%%%%%%%%%%%%%%%%%%%%%%%%%%%%%%%%%%%%%%%%%%%%%%%%%

\begin{figure*}
\centering
\includegraphics[width=0.95\linewidth]{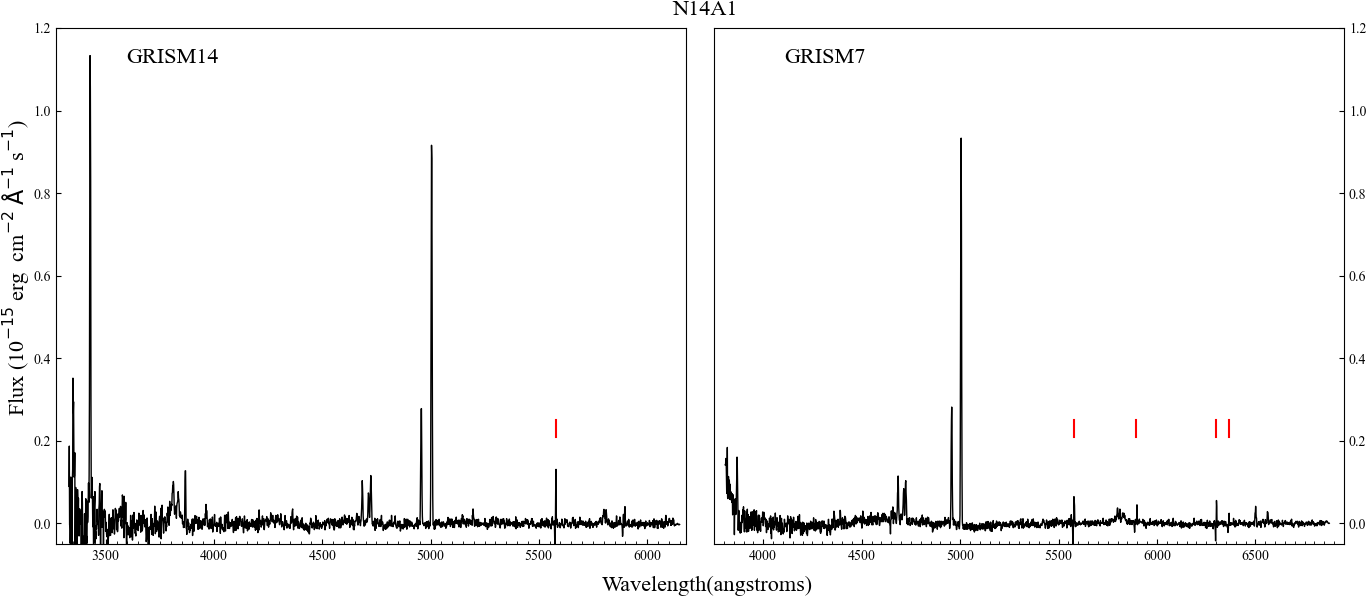}
\\
\includegraphics[width=0.95\linewidth]{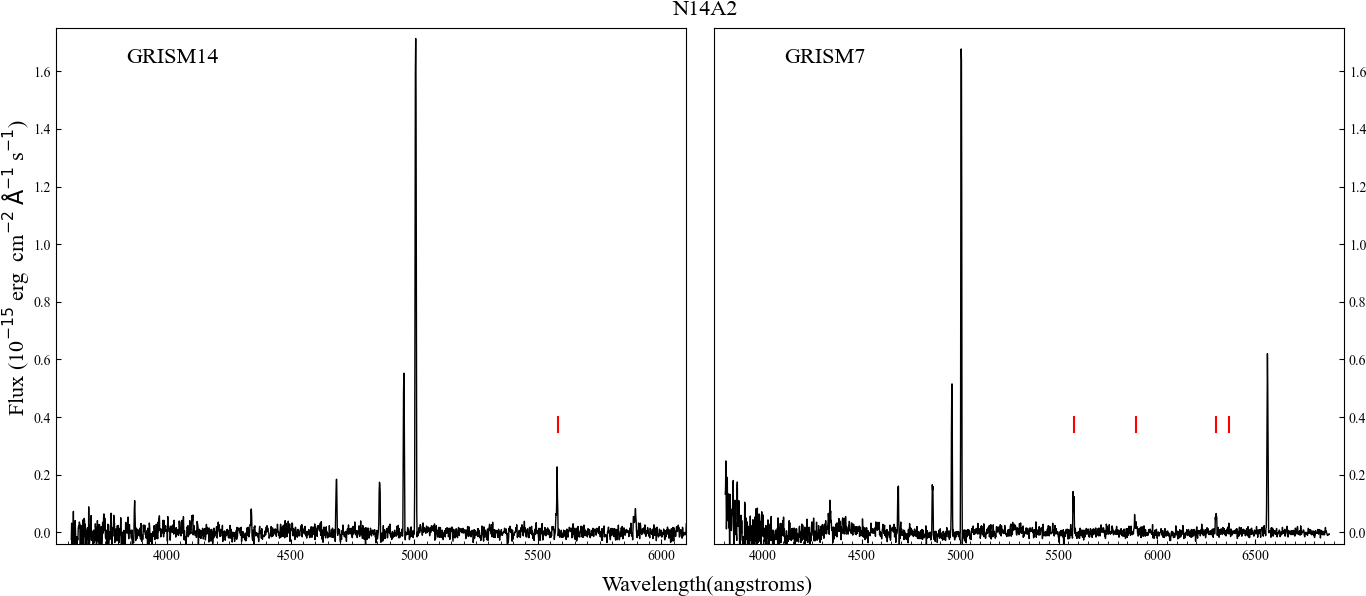}
\\
\includegraphics[width=0.95\linewidth]{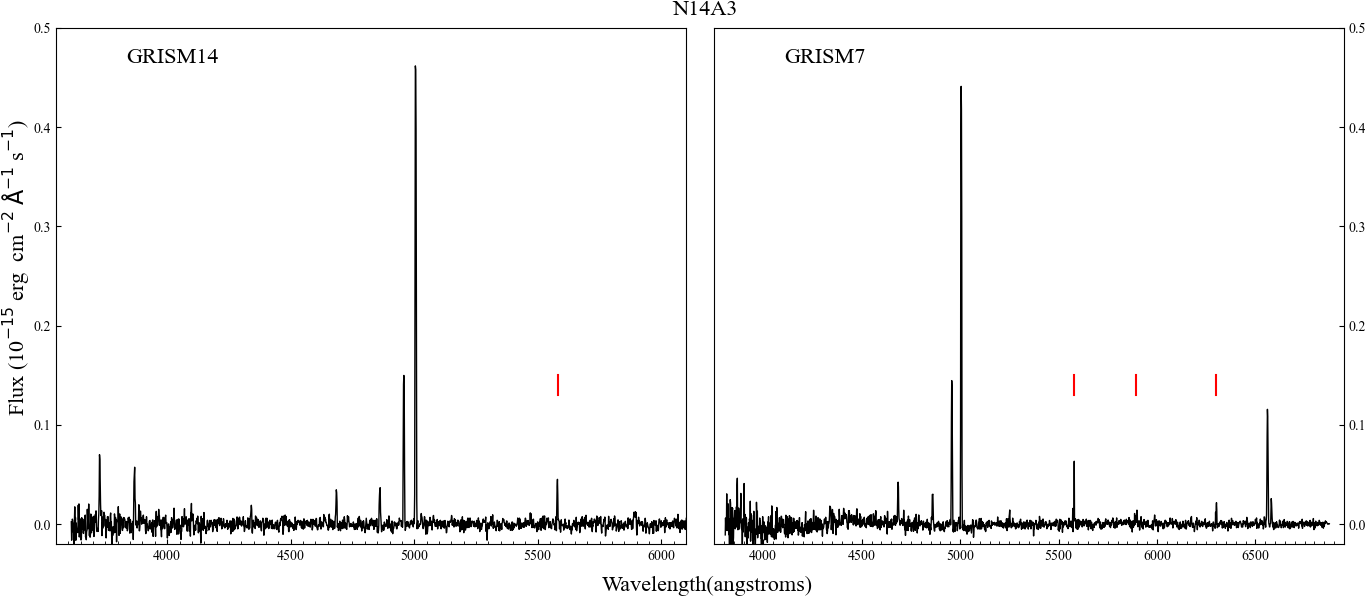}
\caption{
Spectra of the apertures extracted from the 2014.55 observations at NOT. 
The sky emission lines are marked by red ticks to avoid confusing them with nebular emission lines. 
}
\label{fig:A78_spec_n14}
\end{figure*}

\begin{figure*}
\centering
\ContinuedFloat
\includegraphics[width=0.95\linewidth]{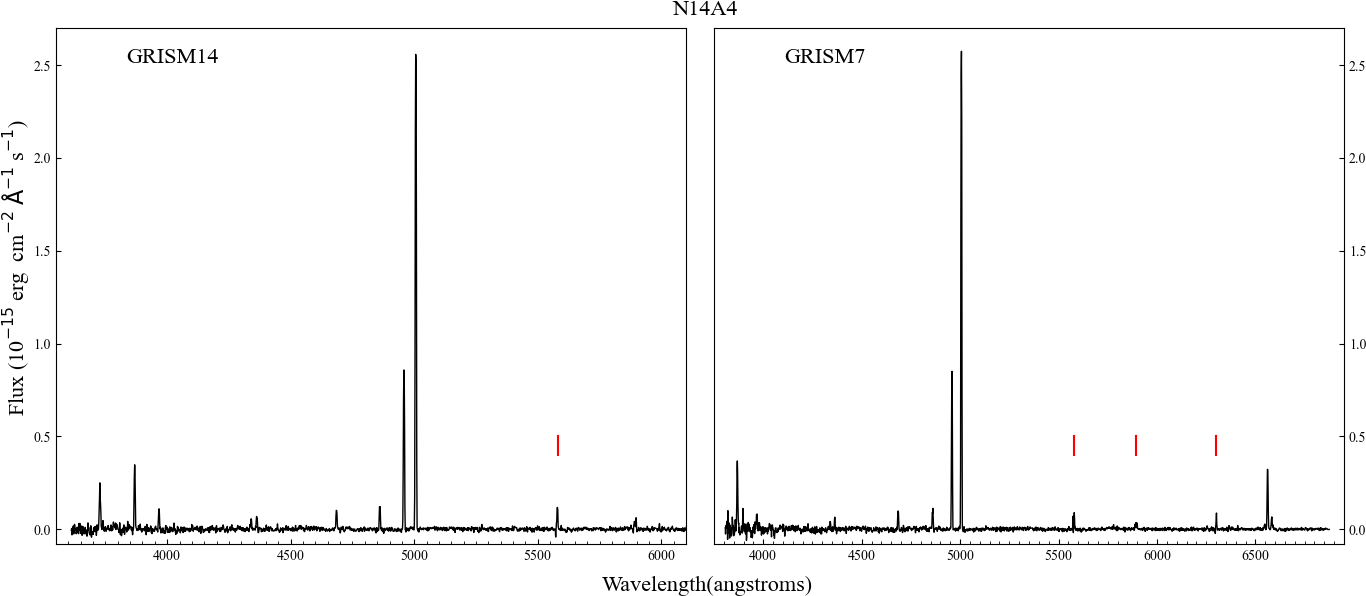}
\\
\includegraphics[width=0.95\linewidth]{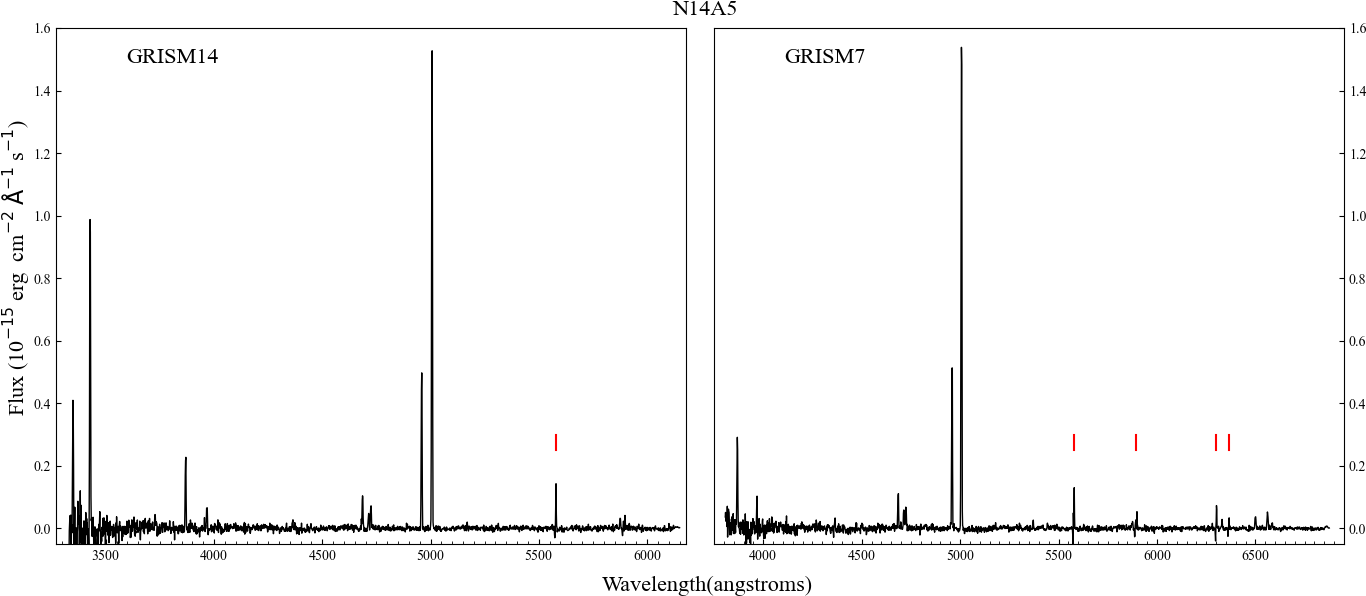}
\\
\includegraphics[width=0.6\linewidth]{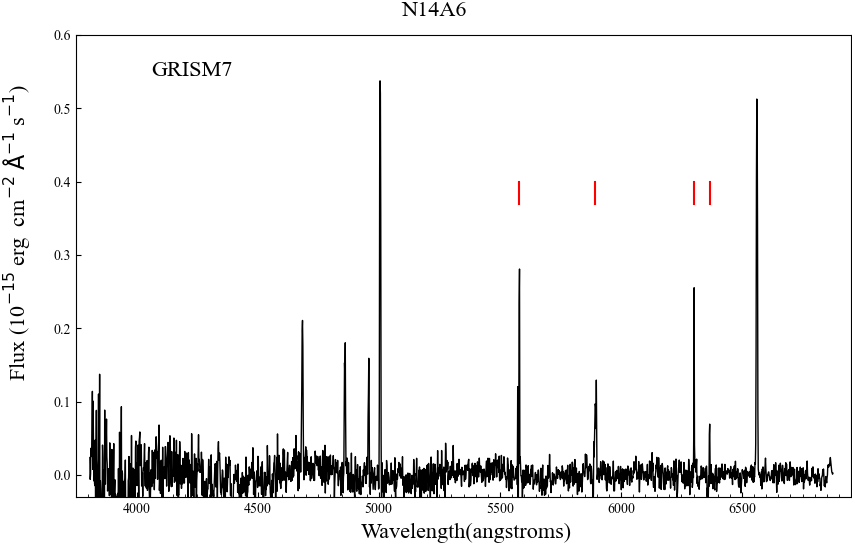}
\caption{
Continued.}
\end{figure*}

\section{Equations system for the iterative removal of the He~II contribution to Balmer lines}
\label{apex:C}

\begin{equation}\label{eqn:system}
\begin{cases}
F(\mathrm{H}\alpha{_c}) &= F\mathrm{(H\alpha)} - F(\mathrm{He}, 6560) \\
F(\mathrm{H}\beta{_c}) &= F(\mathrm{H}\beta) - F(\mathrm{He}, 4860) \\
\mathrm{c(H\beta)} &= \frac{1}{\frac{f6562}{f4861}-1} \cdot \mathrm{log_{10}} \left( \frac{R_{\mathrm{t}} \left( F(\mathrm{H}\alpha) / F(\mathrm{H}\beta) \right)}{R_{\mathrm{o}}\left( F(\mathrm{H}\alpha) / F(\mathrm{H}\beta)\right)} \right)
\end{cases}
\end{equation}

\begin{align*}
F(\mathrm{He\hspace{0.05cm}\textsc{ii}, 6560)} =& F(\mathrm{He} \hspace{0.05cm}\textsc{ii}, 4686)  \cdot 10^{c(\mathrm{H}\beta) \cdot \frac{f(4686)}{f(4861)}} \\
& \cdot R_{\mathrm{t}} \left( \frac{F(\mathrm{He} \hspace{0.05cm}\textsc{ii},6560}{F(\mathrm{He} \hspace{0.05cm}\textsc{ii},4686} \right) \cdot 10^{-c(\mathrm{H}\beta) \cdot \frac{f(6560)}{f(4861)}} \\
F(\mathrm{He\hspace{0.05cm}\textsc{ii}, 4860)} =&  F(\mathrm{He} \hspace{0.05cm}\textsc{ii}, 4686)  \cdot 10^{c(\mathrm{H}\beta) \cdot \frac{f(4686)}{f(4861)}} \\
& \cdot R_{\mathrm{t}} \left( \frac{F(\mathrm{He} \hspace{0.05cm}\textsc{ii},4860}{F(\mathrm{He} \hspace{0.05cm}\textsc{ii},4686} \right) \cdot 10^{-c(\mathrm{H}\beta) \cdot \frac{f(4860)}{f(4861)}}
\end{align*}

%
%(\mathrm{He {\sc ii}}  \cdot 10^{(c(\mathrm{H}\beta)*\mathrm{f}(4686)/\mathrm{f}(4862))) \cdot R_t(\mathrm{H}\alpha/\mathrm{He} {\sc ii} 6560)} \cdot 10^{(-c(\mathrm{H}\beta)*\mathrm{f}(6560)/\mathrm{f}(4862)))}  
 


%%%%%%%%%%%%%%%%%%%%%%%%%%%%%%%%%%%%%%%%%%%%%%%%%%

% Don't change these lines
\bsp	% typesetting comment
\label{lastpage}
\end{document}